\documentclass[12pt,preprint]{aastex}
\shorttitle{PN Velocities in M31}
\shortauthors{Hurley-Keller et al.}

\newcommand{\gsim}{\mathrel{\lower .80ex\hbox{\rlap{$\sim$}\raise.90ex\hbox{$>$} }}}
\newcommand{\lsolar}{$L_{\odot}$}

\newcommand{\mub}{$\mu_B$}

\newcommand{\lon}{[OIII] $\lambda$5007}
\renewcommand{\rq}{$R^{1/4}$}
\newcommand{\vrot}{$v_{rot}$}
\newcommand{\vcirc}{$v_{circ}$}
\newcommand{\vlos}{$v_{los}$}

\newcommand{\sigv}{$\sigma_v$}
\newcommand{\sigz}{$\sigma_z$}
\newcommand{\sigr}{$\sigma_r$}
\newcommand{\sigp}{$\sigma_{\phi}$}
\newcommand{\siglos}{$\sigma_{los}$}
\newcommand{\oa}{$\lambda$5007}
\newcommand{\ob}{$\lambda$4959}
\newcommand{\hb}{H$\beta$}
\newcommand{\voversig}{$v/\sigma$}
\newcommand{\squig}{[OIII]$\lambda5007$/H$\alpha$}
\defcitealias{pvdb}{PvdB}
\defcitealias{nolthenius87}{NF87}
\defcitealias{wk88}{WK88}
\defcitealias{reitzel98}{Reitzel, Guhathakurta, \& Gould 1998}
\defcitealias{reitzel04}{Reitzel, Guhathakurta, \& Rich 2004}
\defcitealias{pat01}{Durrell et al. 2001}
\defcitealias{ferguson00}{Ferguson, Gallagher, \& Wyse 2000}
\defcitealias{ford77}{Ford, Jacoby, \& Jenner 1977}
\defcitealias{ibata94}{Ibata, Gilmore, \& Irwin 1994}
\begin{document}

\title{Planetary Nebulae Kinematics in M31}

\author{Denise Hurley-Keller\altaffilmark{1,2}, Heather L. Morrison\altaffilmark{3} , \& Paul Harding\altaffilmark{1}}

\affil{Astronomy Department, Case Western Reserve University, 10900 Euclid Ave. Cleveland, OH 44106; denise@smaug.astr.cwru.edu, heather@vegemite.astr.cwru.edu, harding@dropbear.astr.cwru.edu}

\altaffiltext{1}{Visiting Astronomer, Kitt Peak National Observatory, National Optical Astronomy Observatory, which is operated by the Association of Universities for Research in Astronomy, Inc. (AURA) under cooperative agreement with the National Science Foundation.}

\altaffiltext{2}{NSF Astronomy \& Astrophysics Postdoctoral Fellow}

\altaffiltext{3}{Cottrell Scholar of Research Corporation and NSF CAREER fellow}

\author{George H. Jacoby}

\affil{WIYN Observatory, P.O. Box 26732,  Tucson, AZ 85726; jacoby@wiyn.org}

\singlespace
\begin{abstract}
\singlespace

We present kinematics of 135 planetary nebulae in M31 from a survey
covering 3.9 deg$^2$ and extending out to 15 kpc from the southwest
major axis and more than 20 kpc along the minor axis.  The majority of
our sample, even well outside the disk, shows significant rotational
support (mean line-of-sight velocity 116 km/s).  We argue that these
PN belong to the outer part of M31's large \rq\ bulge.  Only five PN
have velocities clearly inconsistent with this fast rotating bulge.
All five may belong to tidal streams in M31's outer halo.  One is
projected on the Northern Spur, and is counter-rotating with respect
to the disk there.  Two are projected near the major axis at $X=-10$
kpc and have M32-like velocities; they could be debris from that
galaxy.  The remaining two halo PN are located near the center of the
galaxy and their velocities follow the gradient found by
\citet{ibata04}, implying that these PN could belong to the Southern
Stream. If M31 has a non-rotating, pressure-supported halo, we have
yet to find it, and it must be a very minor component of the galaxy.

\end{abstract}

\keywords{galaxies: kinematics and dynamics --- galaxies: stellar content --- galaxies: individual (M31) --- galaxies: halos}

\newpage
\section{Introduction}
\baselineskip=25pt 
\singlespace

The wealth of information resulting from highly detailed studies of
the old stellar populations of the Galaxy
\citep{els,searle78,gilmore83,edvardsson93} have been a catalyst for
theories of galaxy evolution; as the nearest large galaxy to us, M31
is a proving ground for those theories. M31 offers a variety of
stellar populations in a galaxy of earlier Hubble type, providing
important leverage in testing formation scenarios.  Especially
important are the faint, old populations of the halo, which have
proven difficult to study directly in more distant galaxies
\citep{morrison99,zepf00}.  Halos contain the first stars, and the
ages and abundances of these stars reflect the properties of the
protogalactic fragments that are the building blocks of galaxies
\citep[e.g.][]{oey00}. The kinematics of halo populations tells
us about the dynamical evolution of the galaxy, and provide
constraints on hierarchical formation theories
\citep{helmi99,ivezic00,yanny00,rdp}.

How can we describe M31's old populations?  What can their properties
tell us about how M31 formed and evolved?  The old stellar populations
in the Milky Way have often been used as a guide to these populations
in other disk galaxies because of the richness of spatial and velocity
data for the Milky Way.  In the Milky Way, the metal-poor, old stars
of the halo dominate a few kpc away from the plane and the galactic
center.  The field stars and globular clusters have a power-law
density profile \citep{zinn85,saha85,ivezic00} and pressure-supported
kinematics with little or no rotation \citep[also see][for a recent
summary of halo kinematics in the Milky Way]{beers00}.

This density profile differs from that of M31's spheroid, which has an
\rq\ profile extending unchanged for a remarkably large distance along
the minor axis (20 kpc, \citealt{pvdb}; hereafter PvdB; 30 kpc,
\citealt{pat04}).  Color-magnitude diagram studies of red giants have
shown that, although there is a weak (10--20\%) metal-poor component,
the dominant stellar population at these halo-like distances is about
1 dex more metal-rich than its Milky Way counterpart, or than the M31
globular clusters at the same radius
\citep{mould86b,pat94,couture95,rich96,pat01}. This led
\citet{mould86a} and \citet{freeman96} to suggest that the bulge
dominates even at these very large distances from the galactic center.

The Milky Way's bulge is quite different. It is less luminous and its
inner 1-2 kpc are dominated by a bar, which has a vertical exponential
distribution with scale height of around 350 pc
\citep{weiland94}. Small bulges like the Milky Way's may have formed
by secular processes involving central bars
\citep{pfenniger90,courteau96}, tying the evolution of the bulge to
the disk.  M31's bulge is too large to have formed by this mechanism.

The question of the difference between a large \rq\ bulge with
moderate rotational support and a non-rotating halo is not just one of
taxonomy: the formation processes of these two populations are quite
different. A kinematically hot, non-rotating stellar halo is most
simply formed {\bf after} the aggregation of the galaxy's dark halo
--- this explains the lack of correlation between the angular momenta
of the halo and the disk \citep{freeman96}.  This scenario is
supported by growing evidence that the Milky Way halo was populated,
at least in part, by the accretion and destruction of small satellites
\citep{majewski94,helmi99,ivezic00,yanny00,rdp}. The Sgr dwarf
\citepalias{ibata94} and its tidal streams give a present-day example of
this process.

But the accretion history of the Milky Way may be dramatically
different from that of a galaxy of earlier Hubble type like M31.  An
\rq\ profile is a result of violent relaxation such as occurs, for
example, in major mergers. We note that both accretion and merging
involve interaction between galaxies, but the mass ratios are very
different: in the merger which forms an \rq\ law profile, the mass
ratio is at most a few to one, while the mass ratio for the accretion
of small satellites to form a hot halo is more typically 100:1 or
more.  As a result, major mergers have a strong effect on the galaxy
disk while the accretions have almost none.

Does M31 actually have a kinematically hot halo? If so, how much
luminosity does it contribute?  Accurate kinematic data for halo
objects are needed to {\bf quantify} the relative importance of this
process in populating the outer regions of M31.  However, the
difficulty in isolating the halo from the bulge and in obtaining
accurate velocities for halo populations has precluded a definitive
answer to these questions.  Spectroscopy of single field stars in M31
is very challenging, even for giant-branch tip stars using Keck, as
samples are dominated by contamination by foreground stars from the
Milky Way (\citetalias{reitzel98}; \citealt{reitzel02}).

Globular clusters have been the halo tracer of choice thus far, but
present several difficulties.  Cluster velocities derive from
absorption spectra, and accurate velocities (errors $<$ 20 km/s) are
only now becoming available (\citealt{perrett02}; P. Seitzer 2002,
private communication).  Thus, early studies of M31 cluster kinematics
concluded that its halo had little rotation \citep{huchra91,huchra93},
while more recent studies find greater rotational support
\citep{perrett02}.  In the Milky Way, the field stars and globular
clusters share a similar metallicity distribution
\citep{zinn85,laird88,ryan91} and "hot" kinematics
\citep{norris86,beers00}.  However, in M31, the abundance distribution
of the clusters and the field stars at similar radii are quite
different; the clusters are more metal-poor in the mean than the field
stars (\citetalias{pat01}; \citealt{barmby00}).  Because cluster
metallicities differ, cluster kinematics also may not be entirely
representative of the stellar halo properties.

Planetary nebulae (PN), in contrast, trace stellar populations with
ages $\sim$1--10 Gyr and readily give accurate velocities with
4m-class telescopes, because almost all their energy output is
concentrated in a few emission lines.  The techniques for detecting PN
have been well-documented by G. Jacoby and collaborators
\citep{jacoby89}.  PN have been used extensively as kinematic tracers
in normal ellipticals \citep{hui95,arnaboldi96,mendez01}, and in the
luminous members of the Local Group \citep{nolthenius86,nolthenius87}.
\citet[hereafter NF87]{nolthenius87} presented kinematics of 37 M31
planetary nebulae, with distances along the major axis of up to 40 kpc
and distances from the major axis of up to 17 kpc.  Although it is
possible to identify the signatures of disk and bulge in their data
and find a few objects with clear halo kinematics, the small size of
their sample precludes strong conclusions.

Here we present the results of a kinematic survey of 135 PN in one
quadrant of M31's outer bulge/halo.  The survey extends to 20 kpc
along the minor axis, well into the canonical halo regions of M31, and
20 kpc along the southwest major axis.  We have used both the spatial
and kinematic information for the PN to distinguish the bulge
population from the halo population.  With this larger number of PN
and the accurate velocities (errors of 5--10 km/s) resulting from
the follow-up Hydra spectroscopy, we have a clearer picture of M31's
spheroid.  Using the Milky Way as a guide, we expected to find only a
small degree of rotational support in the populations far from the
major axis, similar to the Milky Way's stellar halo. Instead, we find
that the bulge dominates out to the limits of our survey, and that
very few objects belong to a non-rotating, pressure-supported halo
like the Milky Way's.

In Section 2 we describe the survey which identified our PNe and the
spectroscopic follow-up to obtain velocities. In Section 3 we present
the PN kinematics and discuss the spatial and kinematic signatures of
the thin disk, halo and bulge in M31 with relation to these
results, concluding that a thin disk and a dominant bulge best
explain our data. In Section 4, we interpret the kinematics in the
context of formation scenarios for these populations. 

\section{Observations}

\subsection{PN Detection and Photometry}

Our PN detection relies on the fact that PN have a very strong OIII
emission line at \oa and relatively little continuum emission.
Distant PN can therefore be detected via narrow-band imaging at \oa.
Exposures at a nearby but offset wavelength can be used to
discriminate PN from stars.

Imaging of M31 was carried out at the CWRU Burrell Schmidt on the
nights of 23-30 September 2000.  The camera at the Schmidt has a SITe
2K$\times$4K back-illuminated CCD with 15 micron pixels and a pixel
scale of $\sim 1.45$\arcsec/pixel. It covers a 40\arcmin (E-W)
$\times$ 80\arcmin (N-S) field.  The readnoise during the Schmidt
imaging run was $12 e^-$ and the gain was $1.8 e^-$/ADU.  The seeing
(2--2.5\arcsec) was typical for the Schmidt.

\begin{figure}
\includegraphics[scale=0.8]{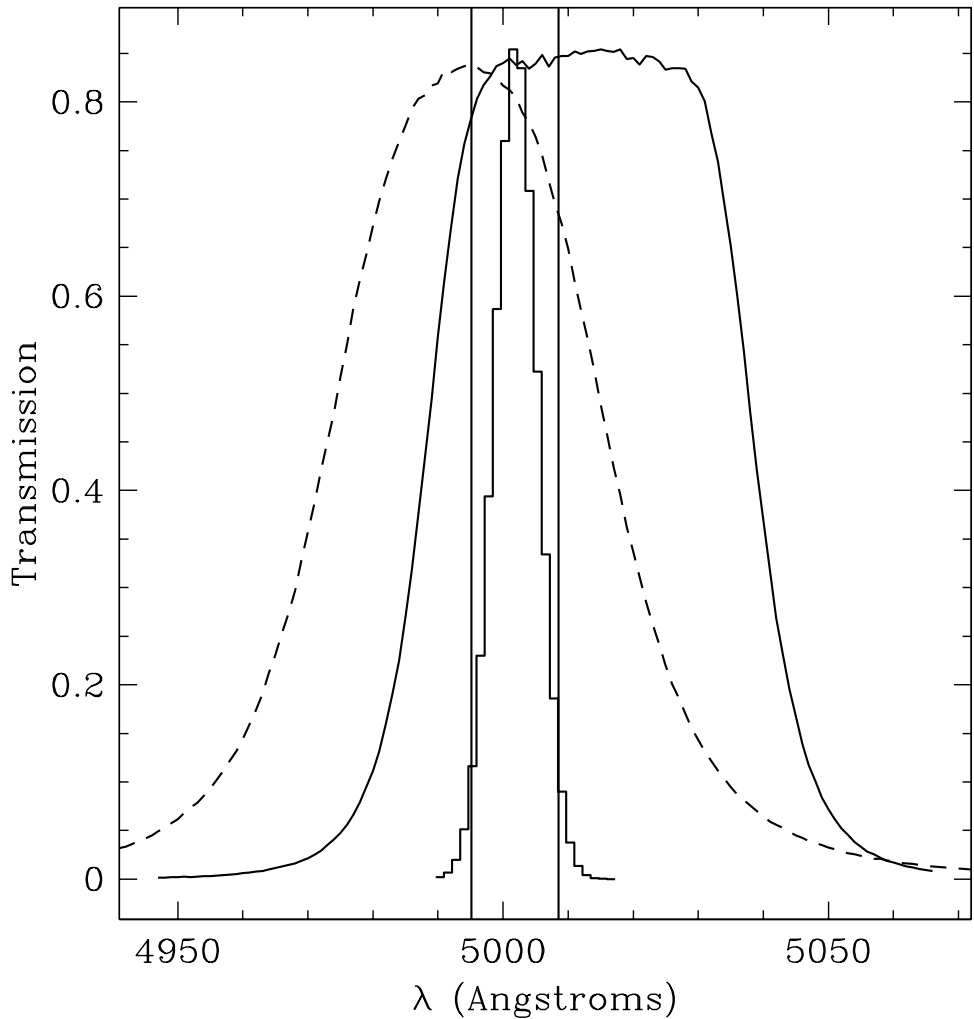}
\renewcommand{\baselinestretch}{0.5}
\caption{Filter transmission curves for this survey.  Solid line:
curve for KP1590, used for Fields 1 and 2.  Dashed line: curve for
KP1467, used for Field 3.  Both curves have been shifted 13 \AA\
blueward to account for the f/3.5 beam of the Schmidt and the
observing temperature.  The histogram is a Gaussian distribution with
mean --300 and $\sigma$ = 200 km/s representing the expected M31 halo
velocity distribution.  The vertical lines show the location of
velocities $\pm$ 400 km/s from M31's systemic velocity.
\label{filters}}
\end{figure}

The choice of filter depends on several factors, and some issues
regarding filters are discussed in \citet{jacoby92}.  The \lon\ filter
should be narrow enough to minimize the sky contribution, yet wide
enough to accommodate the potentially large range of velocities.  It
should also be well-matched to the observed wavelength of the \lon\
emission line.  M31's systemic velocity is --300 km/s
\citep{devauc91}, so that the \lon\ line is shifted blueward to
5002\AA.  A blueward shift of 13 \AA\ in the filter response results
from the Burrell Schmidt's f/3.5 beam (this accounts for a shift of
$\sim$10\AA), and the temperature difference between the lab where the
filter response curve published by NOAO was measured and the observing
conditions ($\sim 3$\AA).  For Fields 1 and 2, we used filter KP1590
with $\lambda_c$ = 5022 \AA\ and for Field 3, filter KP1467 with
$\lambda_c$ = 5008 \AA.

The filter transmission for a velocity of $\pm$400 km/s with respect
to M31's systemic velocity (the likely worst case) is $\gsim$ 70\% in
all cases (Figure \ref{filters}).  We note that the filter used for
Field 3 has a 15\% gradient across the velocities of interest.
Detection of disk PN is unaffected, as the field is located on the
approaching side of the galaxy and the emission is blueshifted into
the higher sensitivity region of the transmission curve.  Also, the
gradient across disk velocities will be negligible due to the
relatively low dispersion of disk PN ($<$100 km/s).  Detection of halo
objects may have been slightly biased against those with velocities
more than 200 km/s greater than the systemic velocity in that field.
Although we have not quantified this effect, it would not alter our final
conclusions.

\begin{figure}
\includegraphics[scale=0.8]{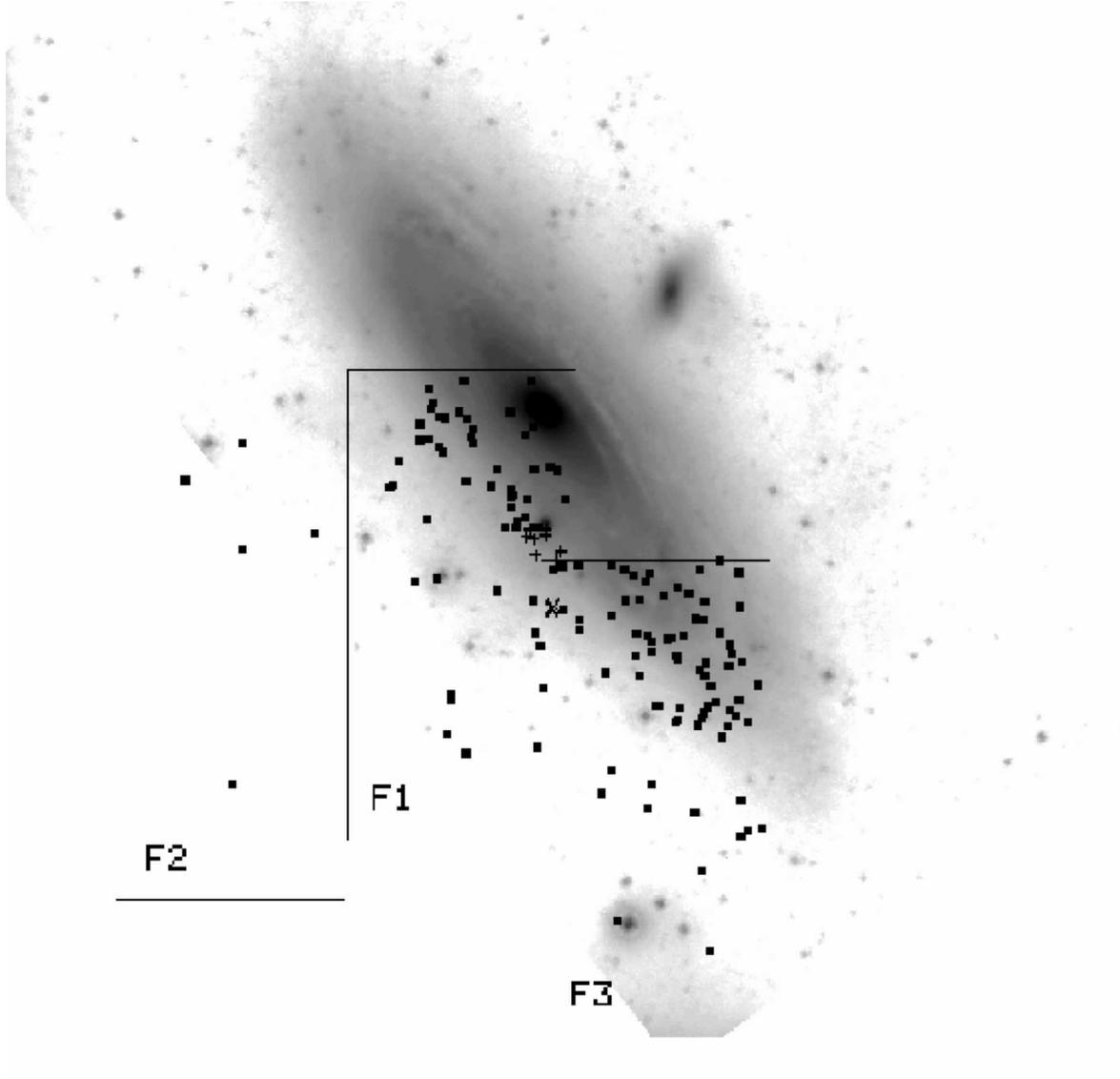}
\renewcommand{\baselinestretch}{0.5}
\caption{Positions of our Schmidt fields and PN in M31.  The R-band
image was kindly provided by Rene Walterbos. North is up, East is
left.  M31's position angle is $37.3\degr$ \citep{devauc58}.  The
SW end of the galaxy is approaching, and the far side is to the SE of
the major axis.  Squares: PN in M31. Pluses: PN in M32.  Xs: HII regions
 in And IV.  
\label{image}} 
\end{figure}

We placed our fields to sample halo PN and to measure their rotation.
Fields 1 and 2 lie along the minor axis of M31, and Field 3 covers a
region at larger radius which includes the major axis.  Figure
\ref{image} shows their placement on an image of M31 kindly provided
by Rene Walterbos.  Field 1 was observed under photometric conditions;
Fields 2 and 3 were observed during periods of thin cirrus and
occasional heavy cloud.  Fortunately, these observations are still
usable because the focus of this project is the PN velocities rather
than the photometry.  We followed standard image reduction procedures.
The images were bias-subtracted and flatfielded, and then registered
using bright stars in the field.  We averaged them using
IRAF\footnote{IRAF is distributed by the National Optical Astronomy
Observatories, which is operated by the Association of Universities
for Research in Astronomy, Inc. (AURA) under cooperative agreement
with the National Science Foundation.} {\it imcombine} with {\it
ccdclip} rejection.

Because we were primarily interested in the kinematics of the halo and
populations, we wanted to exclude HII regions from our sample as much
as possible, as they belong to the young, thin disk.  The geometry
discussed in Section 3.1.1 shows that HII regions confined to the disk
should be rare more than about 3 kpc from the major axis.  Therefore,
we limited our selection of PN from the Schmidt data to those more
than 3 kpc from the major axis near the center of the galaxy.  At
larger radius, where the background light from the disk was fainter,
it was easier to distinguish point-like PN from more extended HII
regions and we allowed the sample to extend to the major axis.
Jacoby, Ford \& Hui kindly shared positions for some of the objects
from their earlier survey of the M31 disk and bulge \citep{hui94}.
During the follow-up spectroscopy run, we assigned spare fibers to
these objects, some of which are nearer the center than the Schmidt
sample.

To identify PN candidates, we generally followed the procedure
outlined by other PN studies such as \citet{ciardullo89}.  For our
project, however, it was not possible to perform photometry on a
difference of the on-band image and the appropriately scaled off-band
image.  This is because the PSF varies across the field of view of the
Schmidt -- due to the curved focal plane -- and depends sensitively on
the seeing and focus.  Large residuals on the difference image would
lead to many false detections.  Instead, we used DoPHOT \citep{dophot}
on the combined on- and off-band images separately and matched the
resulting object lists.  We then matched the position of objects in
the on-band image with objects in the off-band image within a 2 pixel
radius.  Objects in the on-band image which had no counterpart in the
off-band object list were classified as PN candidates.

\begin{figure}
\includegraphics[scale=0.8]{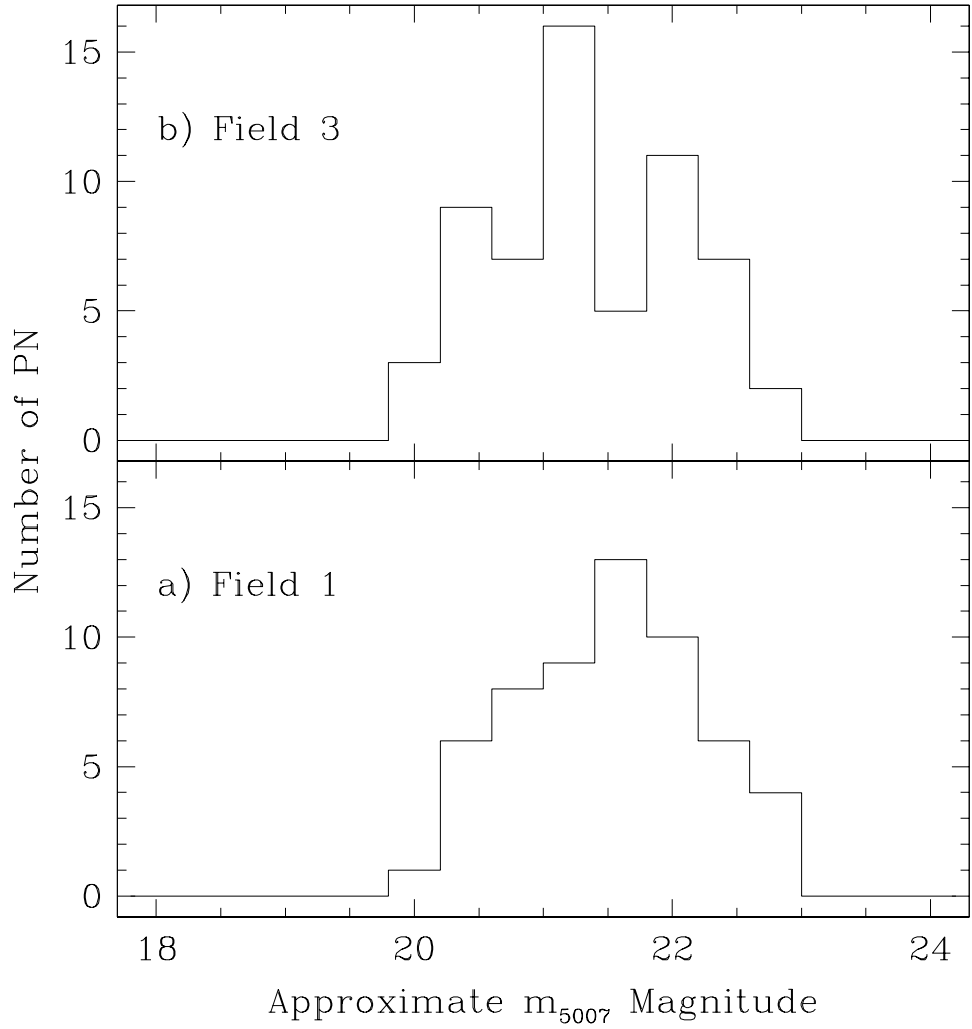}
\renewcommand{\baselinestretch}{0.5}
\caption{The estimated PN luminosity function for Fields 1 (panel a)
and 3 (panel b).  The X axis is the instrumental \lon\ distribution
shifted so that the bright edge of the distribution is at $m_{5007}$ =
--4.5.
\label{PNLF}}
\end{figure}

All candidates were visually verified.  To be included in our list of
spectroscopic targets, a point-source in the on-band image should have
no discernible flux at the same position in the off band image. Faint
candidates were also required to have at least some signal present in
each of the four exposures taken under the most transparent
conditions.  We ranked the PN candidates by the degree to which they
satisfied all the criteria.  As we will discuss below, $\sim88\%$ of
our first priority PN candidates were confirmed spectroscopically,
validating our selection criteria.  Figure \ref{PNLF} shows the PN
luminosity functions for confirmed PN in Fields 1 and 3.  The rough
calibration is based only on assigning the brightest PN to $M_{5007}$
= --4.5 \citep{ciardullo89}.  We reached basically to equal depths of
1.5--2 magnitudes down the PN luminosity function in these two fields.
There were only 5 objects in Field 2, and their photometry is
consistent with the other two fields.

\subsection{Astrometry}

The WIYN\footnote{The WIYN Observatory is a joint facility of the
University of Wisconsin-Madison, Indiana University, Yale University,
and the National Optical Astronomy Observatory.} Hydra fiber
spectroscopy of these candidates requires positions accurate to better
than 1\arcsec.  We identified astrometric standards from the USNO-A
2.0 astrometric catalog \citep{monet98} in each of our three fields.
From these reference stars, we computed the transformation between the
DoPHOT coordinates and the celestial coordinate system using the {\it
tfinder} and {\it ccmap} packages in IRAF.  With $\gsim1000$
reference stars in each field and the high accuracy of the catalog
positions, an accurate transformation was possible even over the large
field of the Schmidt.  The rms error in the calculated positions was
less than 0.3 \arcsec\ in all fields.

\subsection{Spectroscopy} 

The spectroscopic observations were obtained with the WIYN Hydra fiber
spectrograph on the night of 7 September 2001.  The conditions were
clear with good seeing, and we were able to obtain spectra for all of
our candidates.  For the instrument configuration, we used the red
fiber set with the 600 lines/mm grating in first order.  This provided
a wavelength range from 4800 to 6800 \AA, a dispersion of 1.4 \AA\ per
pixel, and a resolution of 2.9 \AA.

The three Schmidt fields were divided into 6 separate Hydra fiber
setups.  For each setup, we took a series of four 600s exposures and a
Copper-Argon (Cu-Ar) lamp exposure.  At the beginning of the night, we
took a series of dome lamp flatfield exposures at a neutral position.
More accurate flatfielding was unnecessary, as the PN in our sample
are relatively bright and we are only interested in velocities.  The
spectra were overscan subtracted, trimmed and bias subtracted via
standard IRAF routines.  The individual fiber spectra were extracted
using the fiber apertures determined from the dome flats.  We then
coadded spectra from the individual exposures to make the single deep
spectrum for each object.

\begin{figure}
\includegraphics[scale=0.8]{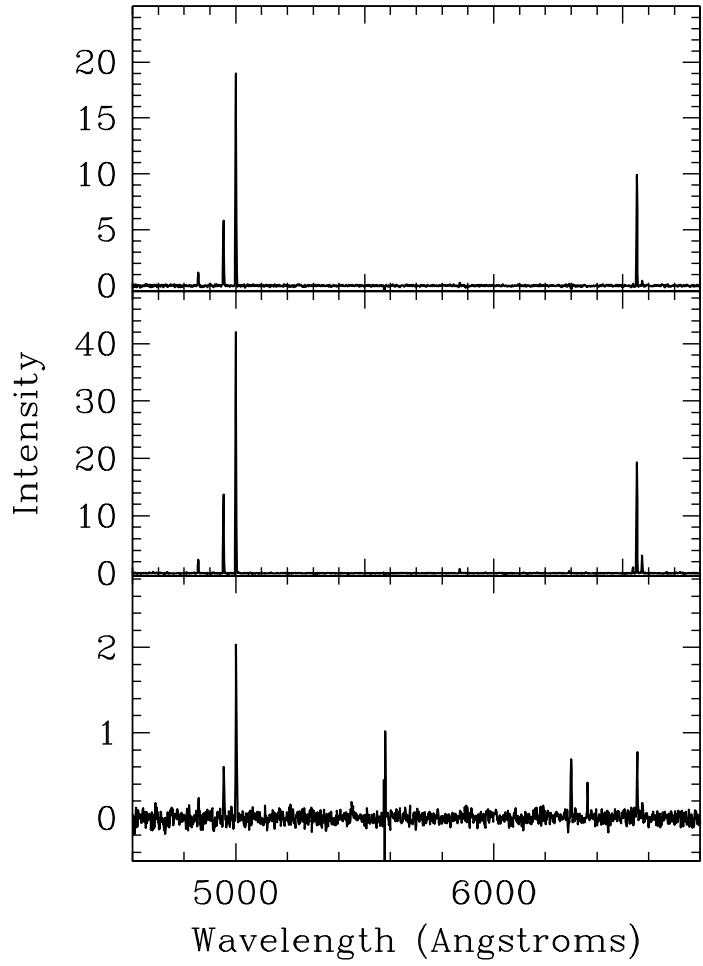}
\renewcommand{\baselinestretch}{0.5} 
\caption{Three typical spectra for our sample, with a range of
signal-to-noise, showing the strongest PN emission lines, [OIII]\ob\
and \oa, and H$\alpha$.  
\label{spectra}} 
\end{figure}

Highly accurate velocities depend upon careful wavelength calibration
of the spectra.  We used the adjacent Cu-Ar lamp exposure for each
setup to wavelength calibrate the individual spectra.  A fifth order
Legendre function was used to fit 60--70 lines in the Cu-Ar spectrum
to determine the spectral solution. The rms error in the fits was
always less than 0.2 \AA, or 10 km/s.  The bench-mounted spectrograph
minimizes the spectral shift.  We confirmed this by cross-correlating
the comparison lamp spectra from different pointings, and found that
the spectra shifted by less than 0.05 \AA\ over the course of the
night.  Figure \ref{spectra} shows three typical spectra from our
sample.

Unlike studies of PN kinematics in distant galaxies where only \lon\
is detected \citep[e.g.][]{hui95,arnaboldi98,mendez01}, the proximity
of M31 PN allows us to easily detect \oa\ and \ob\ in nearly all
of our candidates, and \hb\ in most.  Local Group PN studies which
have used all three lines {\it independently }
(e.g. \citealt{nolthenius87}; \citetalias{ford77}) have found that they
lead to systematically different velocities.  We opt instead to use
all three lines simultaneously, thus improving the accuracy of our
derived radial velocities.

We used the {\it rv} package in IRAF to derive radial velocities using
the cross-correlation of the object spectra with a template.
Initially we created an artificial PN spectrum to serve as the
template spectrum (three lines matched to the instrumental PSF at
\oa, \ob, and \hb), but found that we were able to reduce the
internal velocity errors significantly by using a spectrum created by
combining real PN spectra as a template.  We selected the 8 highest
signal-to-noise spectra from the PN in the first pointing. After
continuum-subtracting them, we shifted them to zero velocity by
fitting a Gaussian to the \oa\ line profile.  The final template
was created by coadding these 8 spectra.  The radial velocities are
derived by calculating the cross-correlation function of the template
and continuum-subtracted PN spectra in the spectral region including
\oa, \ob, and \hb, using {\it fxcor} in IRAF.  In general, we did
not sky subtract the PN spectra as there were no significant sky lines
or stellar absorption lines in this spectral region.  The resulting
radial velocities and errors are listed in Table 1.

\subsection{Velocity Errors} 

All the PN were observed on the same night with the same instrument,
making systematic differences within the PN sample highly unlikely.
We observed 11 PN from our sample in multiple pointings.  The mean
difference in velocity for these PN is --1.2 km/s with a standard
deviation of 4 km/s. These results are consistent with the typical
internal velocity error estimates from {\it fxcor} of 3--7 km/s.  We
estimate an additional 5 km/s of error in the velocities due to
systematics effects.

There have been several earlier studies of emission-line objects in or
near M31, and this allows us to assess the systematic error in our
velocities.  We have spectra for three HII regions in the dwarf galaxy
And IV (objects 3, 4 and 6 from \citetalias{ferguson00}).  The velocities
for objects 4 and 6 agree within 5 km/s (250 $\pm$ 5.2 and 270.8 $\pm$
39.2, respectively compared to 250 and 273 km/s from
\citet{ferguson00}. For object 3, we find a velocity of $233.0 \pm
3.4$ km/s compared to 244 km/s, a difference of 11 km/s.  Our sample
and that of NF have 11 PN in common.  The mean difference $v_{this
paper}-v_{nolthenius87}$ is 10.4 km/s.  This is the same systematic
difference that \citetalias{nolthenius87} note between their
velocities and those of \citet{ford77} for the Galactic PN velocity
standards used in the two studies.  Given the good agreement between
our velocities and those of \citet{ferguson00}, and the disagreement
between the \citetalias{nolthenius87} and \citet{ford77} velocities,
we conclude that there is a systematic offset in the
\citetalias{nolthenius87} velocities.  In the remaining discussion, we
have offset their velocities by +10.4 km/s to agree with ours.

\section{Results}

\subsection{Overview of PN Kinematics}

Although M31 is the primary target of our survey, M32 and the dwarf
galaxy And IV are also located in the survey region.  Membership for
several objects is ambiguous and was decided based on position and
velocity.

The three emission line objects we identified in And IV are HII
regions, resolved in HST images \citep{ferguson00}.  Although PN
commonly have $\xi =$ \squig $\sim 3$, values nearer to 1 are
possible, so that $\xi$ for PN and high-excitation HII regions overlap
significantly. This ambiguity prevents distinguishing PN from HII
regions solely with our spectra; such a distinction requires nebular
modelling which relies on faint lines beyond our wavelength range.
The inclusion of the And IV HII regions in our sample raises the
question of substantial numbers of M31 HII regions in the sample.
However, this is unlikely for several reasons.  The And IV HII region
spectra have $\xi\sim$1/3, much less than the rest of our sample.  And
IV is significantly more distant than M31 ($>$5 Mpc;
\citealt{ferguson00}).  Bright M31 HII regions are more likely to be
resolved, and we considered only point sources as candidates.  Also,
though the value of $\xi$ ranged from 1--3 (excluding the And IV HII
regions), objects with values as low as 1 are rare in our sample
($<$10\%).  With this in mind, we refer to the emission line objects
in M31 as PN throughout the rest of the discussion.

M32 PN are clumped in position and in velocity near M32's radial
velocity of --197 km/s \citep{huchra99}.  For M32, \sigv\ $\sim$ 50 km/s and
the tidal radius is $\sim 2$ kpc \citep{mateo98}.  Nine PN in the
sample have velocities within 50 km/s of M32's systemic velocity and
are located within 1 kpc of M32's center.  PN membership is summarized
in Tables 1, 2 and 3. Each table lists the positions and heliocentric
radial velocities for the emission line objects in M31, M32 and And
IV, respectively.  We have excluded M32 and And IV objects from the
analysis that follows.

\begin{figure}
\includegraphics[scale=0.8]{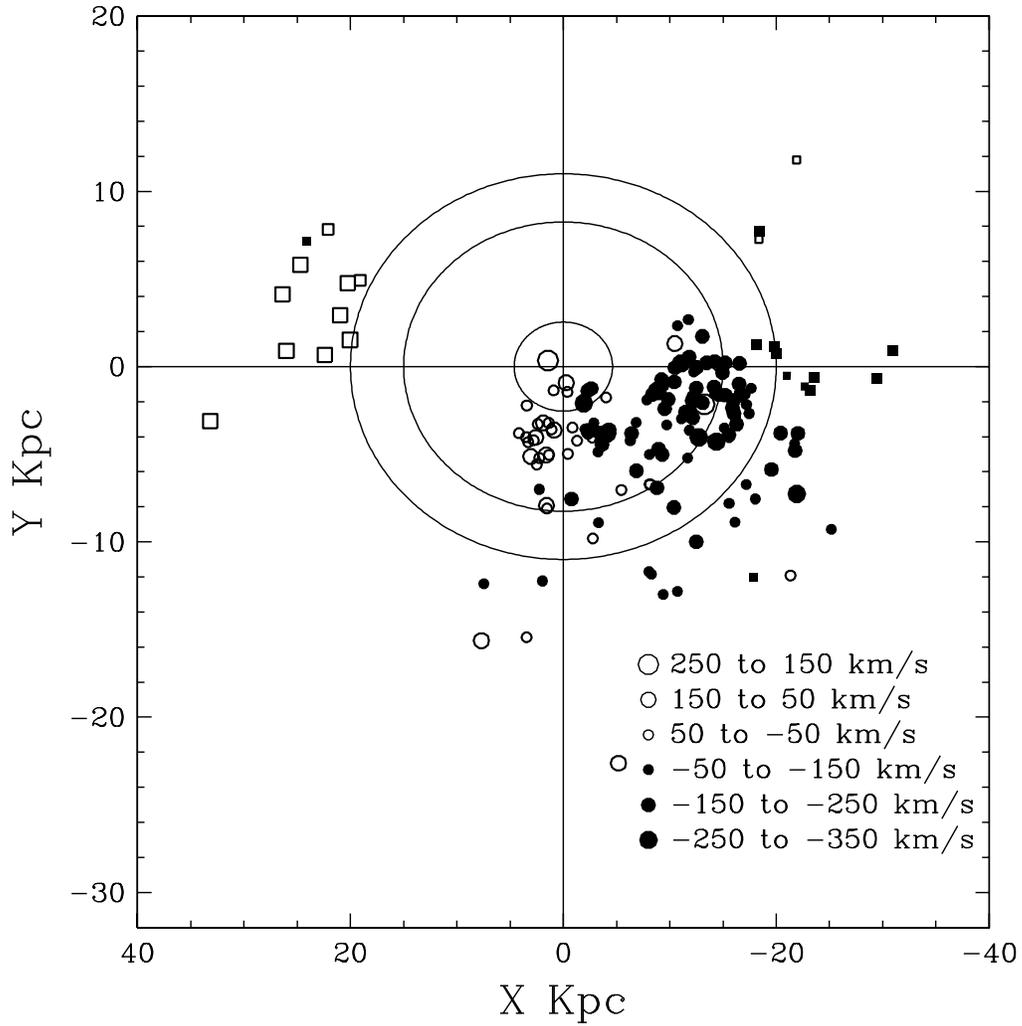}
\renewcommand{\baselinestretch}{0.5}
\caption{Plot showing the positions of PN in M31.  The major axis has
been rotated through the position angle so that it is horizontal in
the figure.  Symbol size corresponds to velocity.  Circles are PN
velocities from this study, and squares are those from
\citetalias{nolthenius87}.  Ellipses are bulge isophotes with
b/a=0.55.
\label{velcoded}}
\end{figure}

The positions of the PN relative to the center of M31 are shown in
Figure \ref{velcoded}.  We have assumed a distance of 770 kpc to M31
\citep{freedman90}.  $X$ is distance in the major axis direction, and
$Y$ is the distance in the minor axis direction.  The positions have
been rotated through M31's position angle
\citep[$37.3\degr$][]{devauc58} so that the major axis defines $Y=0$.
All velocities are plotted relative to M31's systemic velocity and
coded by symbol size.  We have enlarged our sample of PN by adding the
objects from \citetalias{nolthenius87} which are not already in our
sample.

\begin{figure}
\includegraphics[scale=0.8]{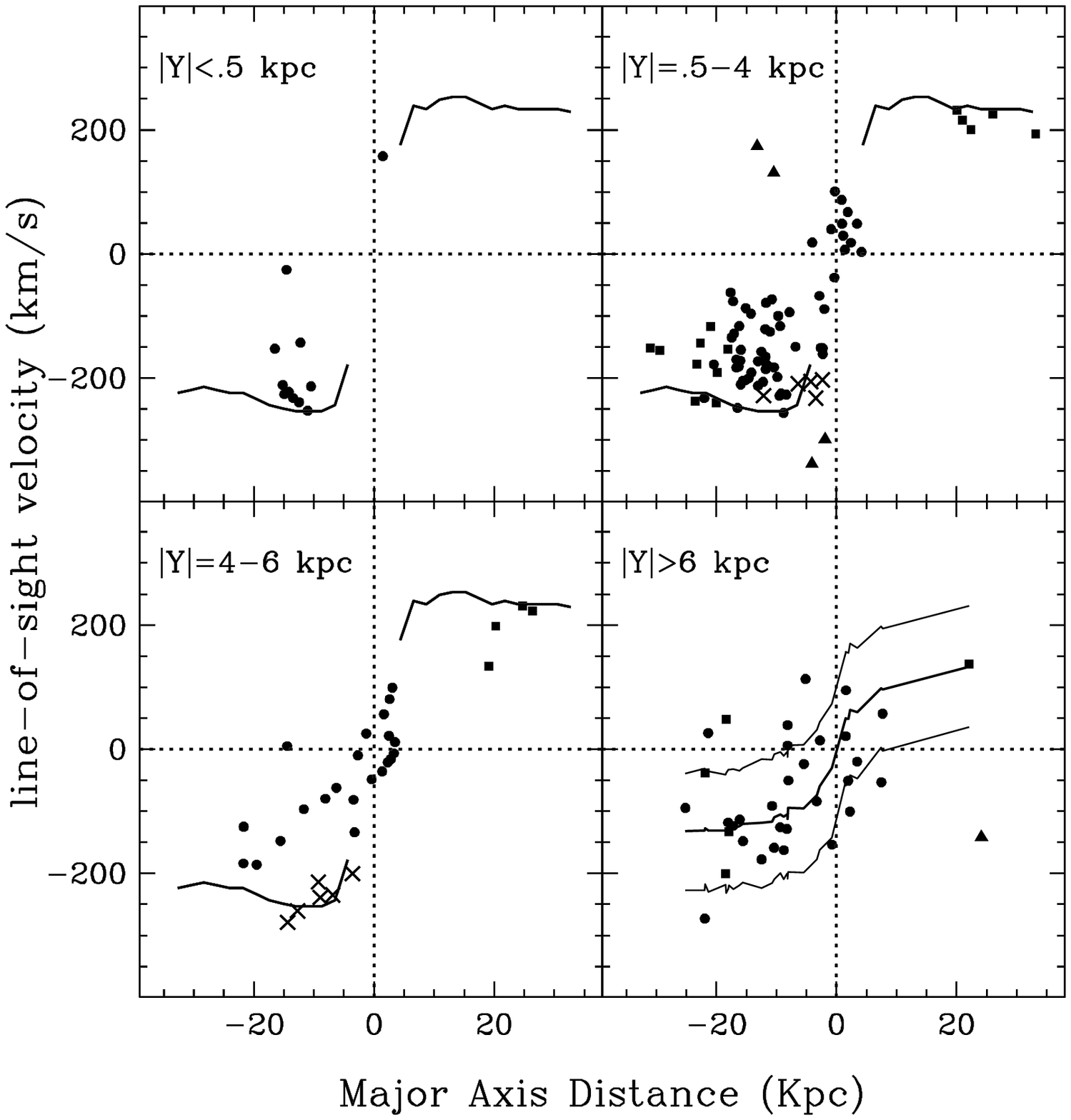}
\renewcommand{\baselinestretch}{0.5}

\caption{Velocity-distance diagram for PN in M31. $X$ is distance
along major axis; $Y$ is the projected distance in the direction of
the minor axis.  Colors for the electronic edition are listed in
parenthesis:  Filled circles (PN from our survey) and squares (PN
from \citetalias{nolthenius87}) are PN with M31 disk or bulge
velocities (black); triangles are halo candidates (red); crosses are
probable And VIII members \citep{morrison03b} (blue).  The rotation
curve is taken from \citet{kent89}.  The solid lines in the lower
right panel represent the range of values which could be expected for
the PN for our best bulge model as discussed in section 3.4.
\label{realveldist}}
\end{figure}

The depth and spatial sampling of the survey is not uniform and must
be taken into account when interpreting the spatial distribution of
the PN.  Only Field 3 intersects the major axis (compare Figure
\ref{image} and Figure \ref{velcoded}).  Fields 1 and 2 had a sizable
gap between them and this can be seen in Figure \ref{velcoded} around
$Y=-12$ kpc.  Fields 1 and 3 are of comparable depth, while Field 2 is
shallower.  Finally, the objects nearest the center of the galaxy,
where we did not attempt to identify PN in the Schmidt images, are PN
from the sample of \citet{hui94} used to fill spare fibers, as
mentioned previously.

The various populations in M31 can be isolated by splitting the sample
in $Y$.  Figure \ref{realveldist} shows plots of distance along the
major axis vs. the line-of-sight velocity for several ranges of $|Y|$.
The disk dominates at $|Y|<0.5$ kpc, where the upper limit on the PN
velocities is the HI rotation curve \citep{kent89}.  A mixture of disk
and bulge objects are seen at $|Y|=0.5-6$ kpc.  Bulge PN near the
minor axis of the galaxy have velocities of --150 to 150 km/s relative
to M31's systemic velocity, consistent with the large bulge central velocity
dispersion of 100--150 km/s \citep{mcelroy83,richstone80}.  Some of
the highest velocity objects in this region (crosses) may be
associated with the newly discovered satellite And VIII
\citep{morrison03b}.  We expect to find bulge and halo objects at
$|Y|>6$ kpc.  We discuss the distinction between these regions further
in Section 3.2.

A striking feature of Figure \ref{realveldist} is the very small
number of PN occupying the ``forbidden'' quadrants for disk objects,
even at large $Y$. This lack of objects in the two non-disk quadrants
is the signature of a rotationally supported population, which clearly
dominates our sample.

Disentangling the various stellar populations in M31 is complicated by
the galaxy's inclination ($77\degr$ from face-on) which causes objects
in the outer disk to appear in projection away from the major axis.
Disk, bulge and halo populations therefore overlap spatially and
kinematically.  The combination of position and velocity data,
however, offers the potential to isolate the different populations.
The allowed velocities for a given PN position will depend on the
kinematic population to which it belongs.  In the discussion that
follows, we present an overview of the expected spatial and
kinematic signatures of the disk and bulge, then use simple
kinematic models to distinguish between them, as well as identify halo
candidates.  We adopt the following definitions: \vlos\ and \siglos\
are the observed, line-of-sight velocity and dispersion, respectively;
for a given population, \vrot\ is the rotation velocity and \sigr\ is
the radial velocity dispersion; \vcirc\ is the circular velocity of
the disk, which reflects the mass distribution, and which we take to
be the HI \vrot, roughly 250 km/s. 

\subsection{M31's Inclined Disk}

\subsubsection{Spatial Extent of the Disk}

The spatial signature of the disk will depend on the size and
thickness of the disk, as well as the degree of warping and the radius
at which the warp begins.  However, we can draw some simple
conclusions from basic geometry.  The radial extent of the optical
stellar disk was estimated by \citetalias{wk88} to be 26 kpc. 
There are now two very deep star count images that can be used to
check this result, from \citet{ferguson02}and \citet{zucker04}. These
images reach to of order 30 V mag/arcsec$^2$, and are thus very
sensitive tests for the extent of the disk. In both images it can be
seen that the brighter portions of the outer disk extend to the
\citetalias{wk88} value, there are some fainter regions which
extend to R=35 kpc and may be associated with tidal streams or with
the outer disk, and that there is no starlight beyond 35 kpc to very
sensitive limits. We have thus assumed $R_{max}$=26 kpc and consider
the effect of a larger $R_{max}$ when relevant in the discussion that
follows.

If the disk were planar and non-warped, its edge would project to a
distance of 5.6 kpc from the major axis because of M31's inclination.
Near the minor axis, the disk and bulge contribute roughly equally to
the luminosity from 2--5 kpc (see Figure 3 of WK88).  We can therefore
expect the bulge and halo PN to dominate beyond $\sim$5 kpc from the
major axis.  These rough geometrical arguments are supported by
observations.  On the side of the major axis covered by our survey,
the outer optical isophotes of the galaxy (\mub\ $\sim26$) extend 4.4
to 6.6 kpc (20--30 arcmin) from the major axis \citepalias{wk88}; HII
regions are found up to 4 kpc \citep{pellet78}.

However, the assumption of a perfectly planar disk is not justified.
Neutral hydrogen studies \citep{roberts75,cram80,brinks84a} have found
two distinct velocity systems for many lines of sight through M31's
disk.  This suggests that the line of sight is crossing the disk
twice.  \citet{brinks84b} interpreted the dual kinematics as a warp
signature.  In their model, the geometry of the galaxy is such that
the line of nodes is nearly aligned with the major axis and the warp
bends over back toward the major axis on the near (NW) side, so that
the warp is nearly edge-on \citep[Figure 2a]{brinks84b}.  The
unwarped HI has a radius of 16 kpc, which would project to a height of
4 kpc from the major axis.  

\citet{brinks84b} introduce a flare of increasing scale height in
their model of the HI warp.  The flare crosses the line of sight to
the major axis and accounts for the two velocity systems there.
Flares are common in the outer HI disks of galaxies
\citep{sancisi79,kulkarni82,olling96}, but the evidence for flared
{\bf stellar} disks is conflicting.  Studies of other galaxies have
generally found a constant stellar scale height at all radii
\citep{vanderkruit81,morrison94}. Recent in situ analyses of Milky Way
stars, however, suggest that its stellar disk may be flared
\citep{lopez02,drimmel03}.  \citet{drimmel03} find that a flare
parameter of 6.6 pc/kpc slightly improves their model fits. However,
this amounts to an increase of less than 0.25 kpc over the entire
length of the disk.  This is is a relatively small increase, and would
be indistinguishable from an unflared disk in our data.

\citet{braun91} analyzed existing HI data and concluded that the warp
model is too simple. He suggested that the data were better explained
by a more complex change in position of the HI disk's midplane,
perhaps caused by gravitational perturbations from M32.  In this
model, the position of the disk midplane varies from 500 pc above the
nominal plane to 1 kpc below it \citep[see][Figure 10]{braun91}.  If
M31's disk had been disturbed by M32's passage, dynamical evolution
may have thickened the stellar disk to an even larger scale height.
However, the cold disk kinematics of some globular clusters rules out
a major interaction, making a disk thickened in this way unlikely
\citep{morrison03a}.  Based on these arguments, we expect that the
disk's luminous influence should have waned entirely by $Y = 5-6$ kpc.

\subsubsection{Kinematic Signature of the Disk}

\begin{figure}
\includegraphics[scale=0.8]{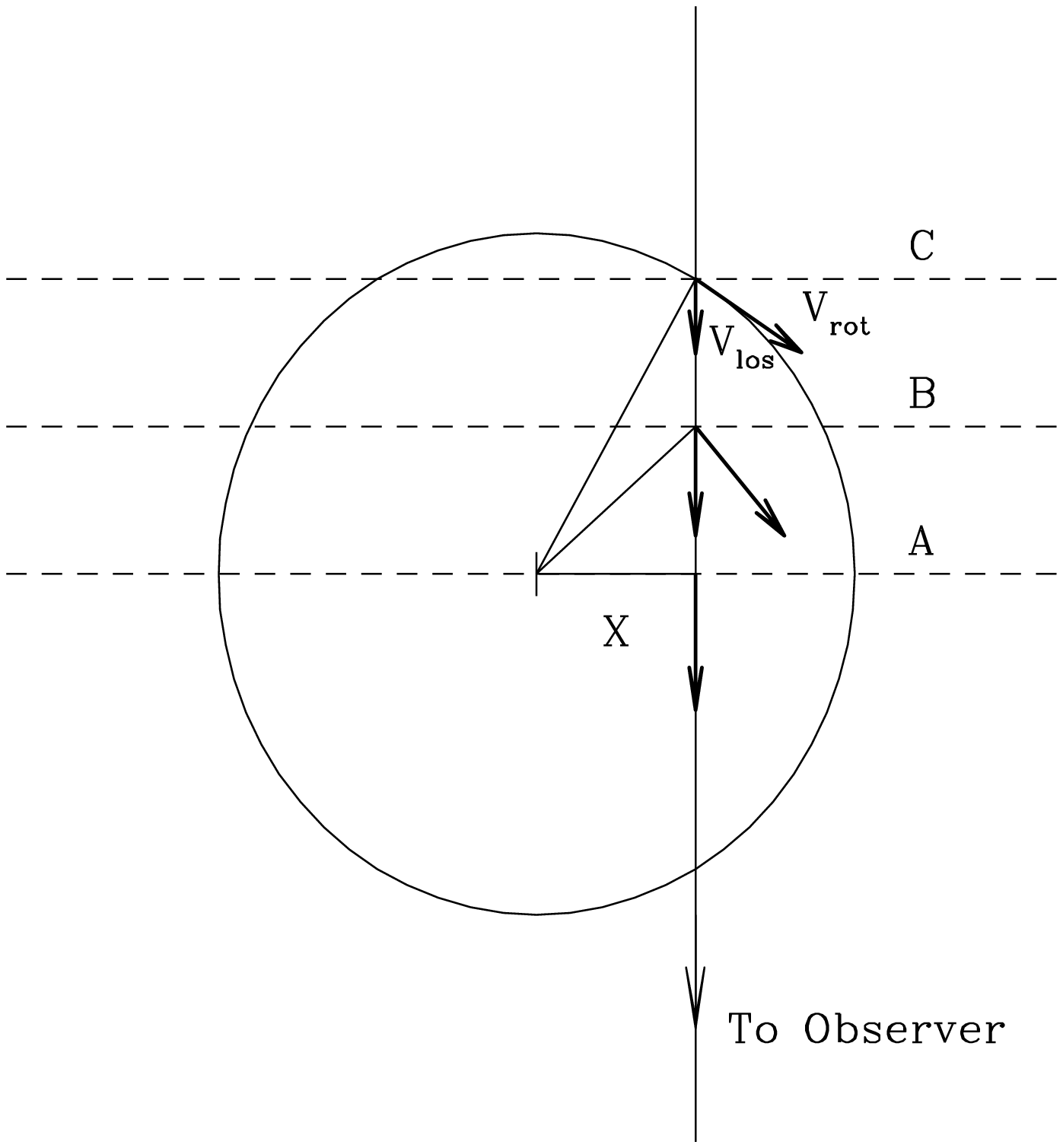}
\renewcommand{\baselinestretch}{0.5}
\caption{Schematic of an edge-on disk, viewed from above.  The
observer is off to the bottom.  \vrot\ is the rotation velocity, and
\vlos\ is the observed radial velocity. The major axis as observed
would lie along line A.  The tangent point occurs at A where \vlos\ =
\vrot\ .
\label{schematic}}
\end{figure}

\begin{figure}
\includegraphics[scale=0.8]{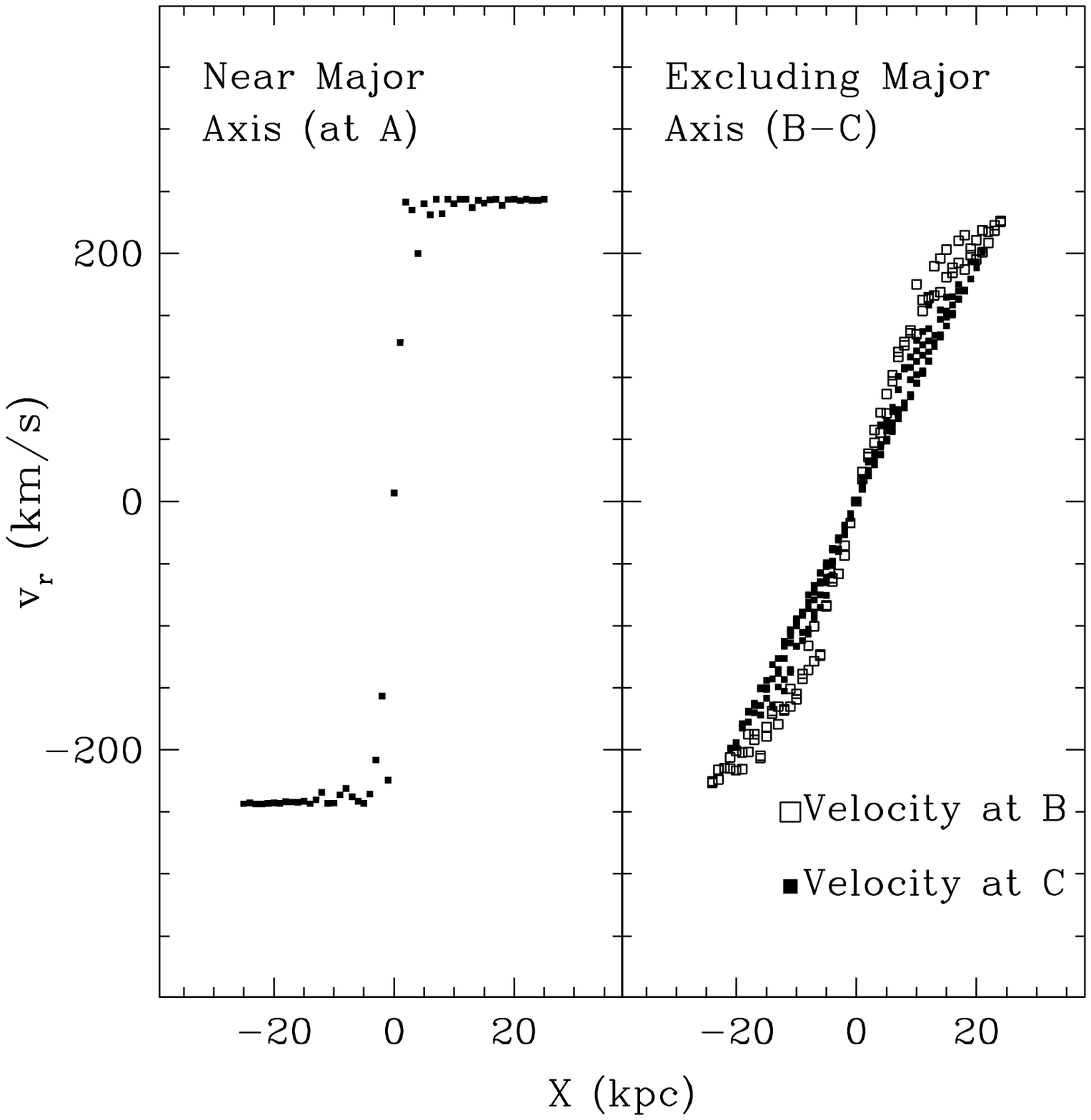}
\renewcommand{\baselinestretch}{0.5}
\caption{Velocities from a model disk, as in Figure \ref{schematic},
but with M31's \vcirc\ and inclination.
\label{modeldisk}}
\end{figure}

The disk's distinctive kinematic signature can be illustrated by
first considering a simple edge-on disk with zero velocity dispersion
and constant circular velocity; a schematic is shown in Figure
\ref{schematic}.  The line of sight through a given major axis
distance, $X$, intersects the disk at different radii resulting in
different projections of the rotation velocity, \vrot, into the
line of sight. At a single value of $X$, any velocity between that at
A and that at C is allowed. The lowest velocity allowed decreases
towards the center of the disk where the line-of-sight velocity,
\vlos, is zero.  Corresponding simulated velocity-distance diagrams
are shown in Figure \ref{modeldisk}.  Disk objects will be found in
only two of the four quadrants of the velocity-distance diagram,
corresponding to the approaching and receding sides of the disk.

Because the disk is inclined rather than exactly edge-on, distance $Y$
from the major axis correlates with position in the disk.  For
example, lines of sight through the major axis intersect the orbital
tangent point (left panel); \vlos\ is the component of \vrot\
projected into the line of sight through the inclination.
Lines of sight through the disk between B and C (right panel) exclude
the major axis points and can have only lower values down to the
lowest allowed by the edge of the disk.  This basic geometry also
holds for a more realistic, hotter disk with some velocity dispersion,
although the velocity dispersion tends to smooth out sharp features.

\subsection{Disk PN Kinematics}

PN near the major axis are those most likely to be disk objects
(Figure \ref{realveldist}, upper left panel).  There are 10 PN with
$|Y|<0.5$ and $|X|>10$.  One object near $X=-13$ kpc has \vlos = --25
km/s, more than 200 km/s from the HI rotation velocity, and probably
belongs to the bulge or halo, so we have excluded it from the disk
sample.  The 9 remaining PN have an average \vlos\ of 215 km/s with
\siglos\ = 39 km/s.\footnote{These values have been corrected for
M31's inclination.  We have dropped the negative sign indicating
rotation toward the observer throughout the remainder of the
paper.}  \siglos\ is dominated by the \sigp\ component of the velocity
ellipsoid on the major axis, implying \sigr\ $\sim$55 km/s.

However, these values of \vrot\ and \sigr\ are very uncertain.  First,
the inclusion of off-axis PN in the average systematically reduces the
measured mean \vlos.  Second, the sample is very small, and the
statistics do not take into account the possible inclusion of bulge
PN.  At this distance along the major axis, the bulge contributes
$\sim15$\% of the total luminosity (WK88), or potentially 1--2 PN of
the 9 in this sample.  Two PN have velocities more than 100 km/s from
the HI rotation and removing those increases the average \vrot\ to 234
km/s, and reduces \sigr\ to 20 km/s.  Third, the disk rotation
velocity is not a secure measurement and is uncertain at the 10-20
km/s level; see the discussion in
\citet{morrison03a}.

With this sample, we cannot distinguish between a thin and thick disk,
and more disk PN are needed \citep[D. Hurley-Keller et al., in
preparation]{merrett03}.  \citet{morrison03a} show that a large
subsample of globular clusters which project on to the disk have thin
disk kinematics.  This suggests there has been no merger with M31
large enough to disrupt its thin disk since the formation of the
clusters and is compelling evidence against a thick disk in M31.

\subsection{Beyond the Disk: Bulge and Halo}

At $|Y|>6$ kpc, the PN kinematics are still dominated by rotation
(Figure \ref{realveldist}, lower right panel).  The mean
\vlos\ of the PN with $|X|>10$ and $|Y|>6$ is 116 km/s, with \siglos\
= 79 km/s.  These PN clearly do not belong to a high velocity
dispersion, non-rotating halo like that of the Milky Way.  To what
population do they belong?

\subsubsection{Overview of Kinematic Models}

To answer this, we constructed simple kinematic models of disk and
bulge populations as described in \citet{morrison03a}.  Briefly, for
each PN, $X$ and $Y$ are known, but the line-of-sight depth is not.
Using a Monte Carlo method, we choose a likely line-of-sight position
based on the density distribution and iterate until a detectable PN
results.  Once we have a line-of-sight depth, we randomly generate an
appropriate velocity for that position in the disk.  The velocity is
then projected into the line of sight.  The result is a predicted
velocity for each PN position given a specific kinematic model.

For the thin and thick disks, we assume an exponential surface
brightness distribution in R and Z;

$L(R,z) = e^{(-R/h_r)}e^{(-z/h_z)}$ 

\noindent where $R$ is the cylindrical radius and $z$ is the height
above the galaxy plane.  

For disk parameters of scale height, scale length and maximum radius,
we have used the Milky Way and other spiral galaxies as templates.  We
adopted a scale height $h_z$ of 0.3 kpc for the thin disk, comparable
to that of the Milky Way thin disk.  For the thick disk, we chose a
scale height of 1 kpc; increasing this to 2 kpc has no effect on our
conclusions.  The thin and thick disk scale length $h_r$ and maximum
radius are fixed at 5.9 kpc and 26 kpc respectively, values taken from
\citetalias{wk88} and scaled for our assumed distance to M31 of 770
kpc.  Although thick disks in other galaxies have larger scale lengths
than the thin disks \citep{morrison99,neeser02}, varying the thick
disk scale length by 20\% had no measurable effect on the models we
discuss in the next section.  Thus far, the few attempts to constrain
the cutoff radius of thick disks suggest that thin and thick disks in
a given galaxy have the same extent \citep{pohlen04}.

At the large values of $X$ under consideration, \vrot\ can be treated
as constant, and we fix \vcirc\ at 250 km/s, a representative value
based on the HI data.  We have adopted a relationship between \vcirc\
and \sigr\ derived from the \citet{vanderkruit84} and \citet{bottema93}
studies of galaxy disks:

\sigr\ $=\onethird v_{circ} e^{-R/2h_r}$

This relationship can be normalized to match the Milky Way thin and
thick disk values of \sigr\ at the solar radius, taken from reviews by
\citet{norris99} and \citet{morrison99} and listed in Table 4.  The 
equation becomes:

\sigr\ $= \sqrt{h_z} v_{circ} e^{-R/2h_r}$

\noindent The velocity ellipsoid follows from the epicycle
approximation \citep[e.g.][]{binney87}:

(\sigr,\sigp,\sigz)$=(\sigma_r,\frac{\sigma_{r}}{\sqrt{2}},\frac{\sigma_{r}}{2})$  

\noindent

For the bulge model, we use an \rq\ surface brightness profile, with
an axial ratio of b/a=0.55 \citepalias{pvdb}, and adopt an isotropic
velocity ellipsoid \large (\sigp\ = \sigz\ = \sigr) \normalsize, with
constant rotation at these large values of $X$.  This simplification
is not unreasonable.  As yet, little is known about the behavior of
bulges at large radii; some bulges exhibit cylindrical rotation, while
others show a decrease in \vrot\ with distance from the midplane
\citep{kormendy82}.  In the current sample, there are too few PN to
test a more detailed model.

We have chosen an axisymmetric model for the bulge, although some
authors suggests that M31's inner bulge is triaxial
\citep{braun91,stark94,berman02}.  We note that the 2MASS image of the
M31 bulge \citep{jarrett03} shows considerably less isophote twisting
than the optical images, suggesting that the disk warp may have
contributed to earlier impressions of a triaxial bulge.  The HI
rotation curve of \citet{braun91} shows a strong central peak which he
attributes to non-circular gas orbits, suggesting a triaxial
potential. However, the CO rotation curve of \citet{loinard99} has a
smaller central peak than the HI curve of \citet{braun91}. Thus, the
bulge may not be strongly triaxial.

\begin{figure}
\includegraphics[scale=0.8]{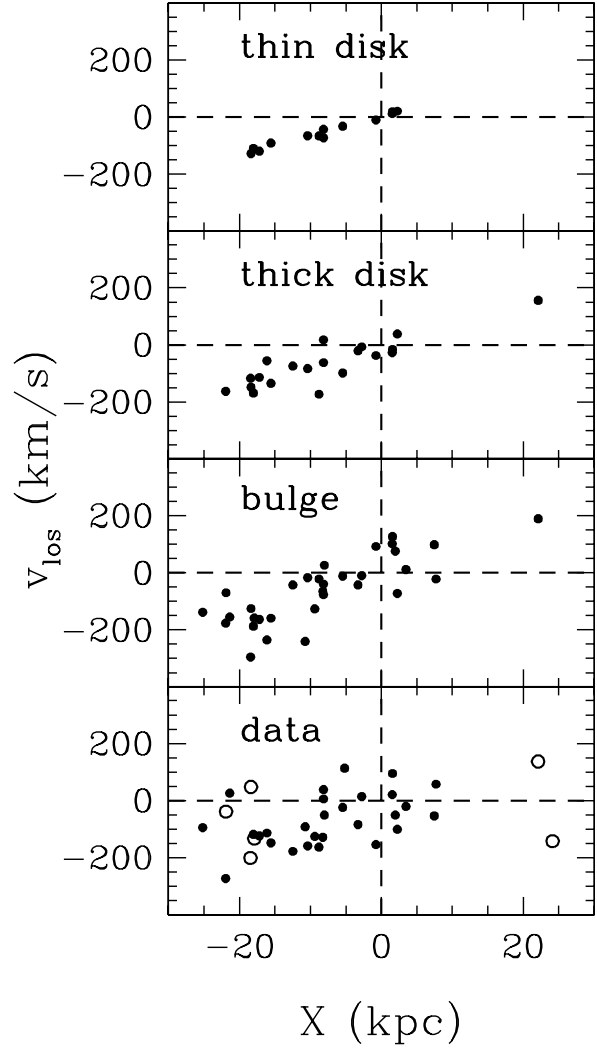}
\renewcommand{\baselinestretch}{0.5} 
\caption{A comparison of thin disk, thick disk, and bulge kinematic
models and the PN kinematics, for $|Y|>6$.  Solid circles are PN from
this survey, and open circles are PN from NF87.  Both the thin and
thick disk kinematics have less dispersion at this height above the
major axis than the data.
\label{thinthick}}
\end{figure}

\subsubsection{Modeling Results}

Representative thin disk, thick disk, and bulge models for the outer
PN are shown with the data for comparison in Figure \ref{thinthick}.
The important properties of these models are summarized in Table 5.
Only a small fraction of the PN have positions which can be fit by the
thin disk model.  Most of the PN (20 of 33) at $|Y|>6$ would deproject
to radii beyond the optical edge of the disk at 25--35 kpc.  Perhaps
more convincingly, the PN velocities are also incompatible with this
model.  Because velocity dispersion falls off with radius in a disk of
constant scale height \citep{vanderkruit84,bottema93} and is low
($\sim20-30$ km/s) for a thin disk at all radii, the disk at this
large radius has a low line-of-sight velocity dispersion.  This
kinematic discrepancy between the model and the data holds regardless
of the maximum radius of the model disk used.

The thick disk model could account for the PN at $|Y|=6-10$ kpc;
however, the velocity dispersion along the minor axis in the model is
significantly less than observed, because PN this far from the major
axis would belong to the outer disk where the dispersion is relatively
low.  By 25 kpc, \sigr\ has fallen to $<40$ km/s in our model of the
thick disk.  The contribution of the thick disk is vanishingly small
by $|Y|=10$ kpc, and the thirteen PN beyond this (see Figure
\ref{velcoded}) are not fit by the thick disk model.  Those thirteen
PN have a significant rotation component, with an average \vlos of
$\sim$70 km/s.

Occam's razor is an appropriate consideration here -- there is no need
to introduce a new stellar population in M31.  The thirteen PN beyond
$|Y|=10$ kpc already represent a rotating population which are not
explained by the thick disk model.  No convincing evidence of a thick
disk in M31 has yet emerged, and the cold kinematics of many M31
clusters suggests that M31 has not experienced a merger massive enough
to form a thick disk since the birth of the globular clusters
\citep{morrison03b}; only satellites with $\sim$10\% of the disk mass
significantly heat the disk \citep{quinn86,walker96}.

\begin{figure}
\includegraphics[scale=0.8]{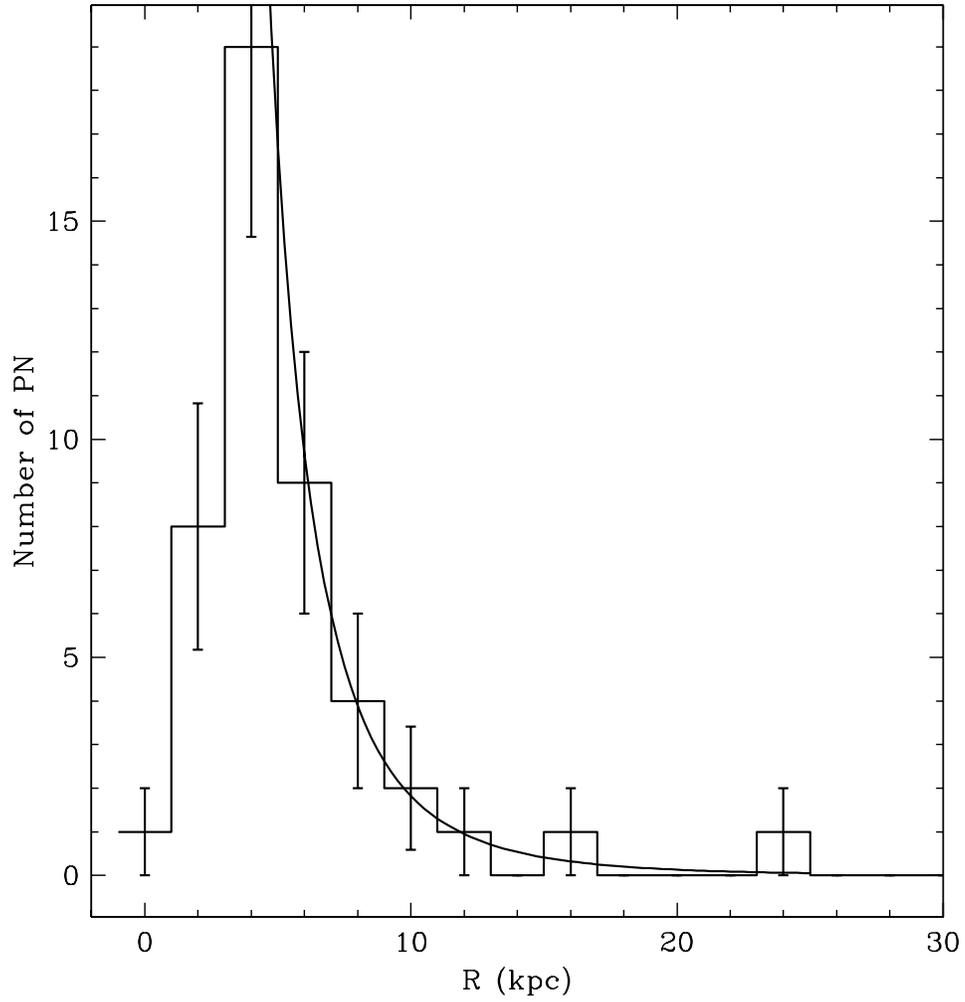}
\renewcommand{\baselinestretch}{0.5}
\caption{The distribution of PN along the minor axis.  PN in the
complete subsample within 3 kpc of the minor axis are assigned an
isophotal distance and binned into 2 kpc bins.  The solid line is the
number of PN expected based on the minor axis surface brightness
profile from \citetalias{pvdb}, normalized to the depth of our survey.
The first three bins are reduced by our exclusion of PN near the center
of the galaxy.
\label{pndist}}
\end{figure}

The bulge, however, offers a ready explanation.  Existing star counts
have already hinted that the bulge extends far into the canonical
halo.  The minor axis profile is well-fit by a single \rq\ profile as
far out as it has been measured (20 kpc; \citetalias{pvdb}).  In
Figure \ref{pndist}, we have binned the complete subsample of Field 1
PN along the minor axis ($|X|<$3 kpc) by isophotal radius, assuming
that the bulge has an axial ratio of 0.55.  We determined the complete
subsample by fitting the known PN luminosity function to the
histogramed data, similar to \citet{ciardullo89}.  At some point, the
number of PN observed falls off from the predicted number of PN and
this determines our completeness limit.  The two PN from Field 2 which
are included in this minor axis sample are the two most distant bins
in the histogram.  The inner three bins are affected by our exclusion
of PN very near the center of the galaxy, as described earlier.

The minor axis PN distribution is well fit by the prediction of the
\citetalias{pvdb} surface brightness profile, assuming the M31 PN
production rate of \citet{ciardullo89} for the upper $\sim 1.5$
magnitudes of the PN luminosity function.  It has been suggested that
a drop in the PN production rate can be caused by both extreme old age
and very low metallicity \citep{ciardullo95,jacoby97,magrini03}, as
expected in a halo population.  We see no dramatic change, however, in
the PN production rate along the minor axis out to at least 12 kpc.
This suggests no sudden decrease in metallicity or age of the dominant
population, but rather continuity from the central bulge out to at
least 12 kpc.

\begin{figure}
\includegraphics[scale=0.8]{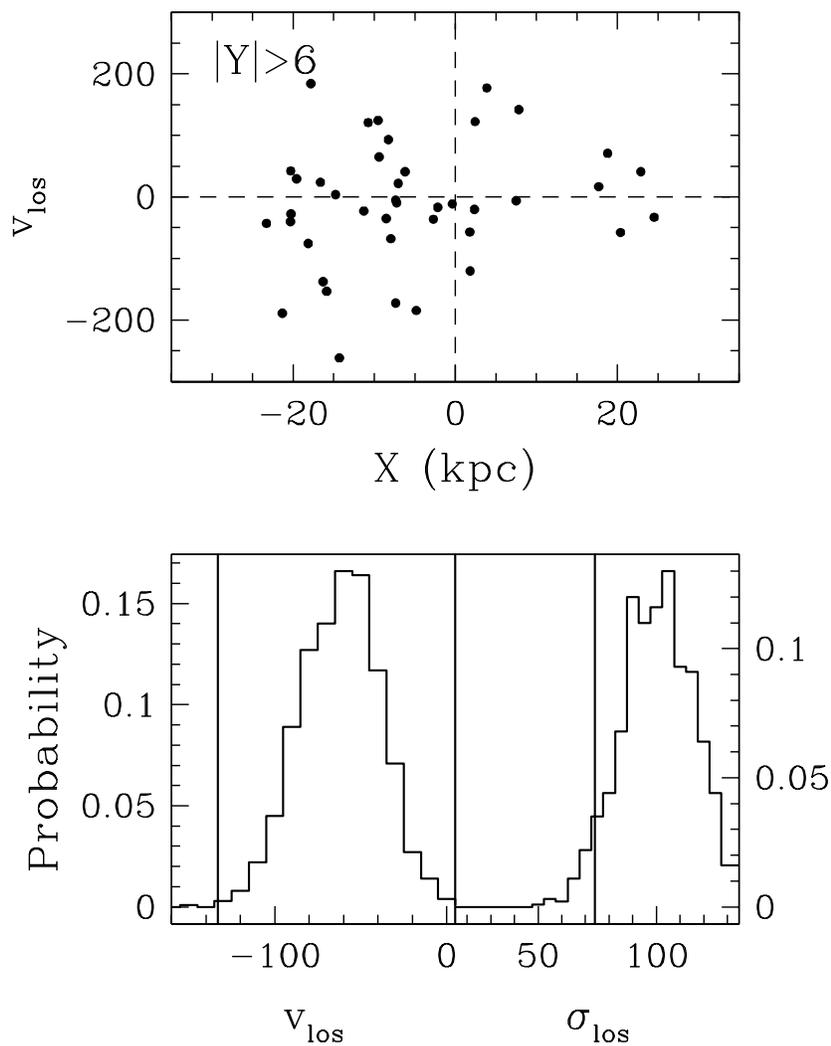}
\renewcommand{\baselinestretch}{0.5}
\caption{Upper panel: One realization of a bulge model with \vrot\ =
80 km/s and \sigr\ = 100 km/s.  Lower panel: The probability
distribution for the mean \vlos\ and for \siglos\ measured from the
positions at $|Y|>6$ and $|X|>10$ for 1000 realizations of the same
bulge model as in the upper panel.  The solid lines represent the
values measured from the data, \vlos\ = 116 km/s and \siglos\ = 79
km/s.  All velocities are in km/s.
\label{zerorot}}
\end{figure}

With this in mind, we feel justified in considering whether the outer
PN velocities are consistent with bulge kinematics.  Like the bulges
of other early-type spirals, which have $v/\sigma_v\sim 1$
\citep{kormendy82,davies83}, M31 has a moderate degree of rotational
support in the inner regions. The central bulge has a large velocity
dispersion (100 km/s) and a rotation curve which rises to 80 km/s by
about 1 kpc \citep{richstone80}. This is significantly less rotation
than seen in the data (116 km/s), and a kinematic model based on the
central bulge characteristics proves incompatible with our data
(Figure \ref{zerorot}).

However, this is not a fair comparison.  The properties of the central
bulge are based on the nuclear observations of \citet{richstone80} and
the stellar absorption data of \citet{mcelroy83}, which only extend to
about 2 kpc along the major axis, less than 1 $R_e$ for the M31 bulge.
These data are on the rising portion of McElroy's rotation curve, and in
fact show an increase in mean velocity with distance from the center
in his Figure 7.  Therefore, the bulge rotation could reasonably
flatten out at 150 to 200 km/s beyond 6 kpc on this side of the
galaxy.

Recent observations of low luminosity ellipticals \citep{rix99} show a
similar increase in the degree of rotational support at large radii
($>1R_e$), with \voversig\ = 2--3 in many cases.  M31's bulge occupies
the same region of the fundamental plane as low luminosity ellipticals
\citep{kormendy85,kormendy87,kormendy99}, and we might therefore
expect that they have common kinematic properties.

The low-luminosity elliptical data are derived from stellar absorption
spectra along the major axis, while our outer PN are far off-axis.  In
order to make a better comparison, we have used our kinematic models
to predict the value of \vrot\ and \sigr\ that we would expect to
observe along the major axis if our PN belong to a bulge with
increasing rotational support. We generated 1000 realizations for each
of a set of models covering a range of possible
\vrot\ and \sigr\ values, and compared the resulting probability
distributions for our variables to our observed values.

\begin{figure}
\includegraphics[scale=0.8]{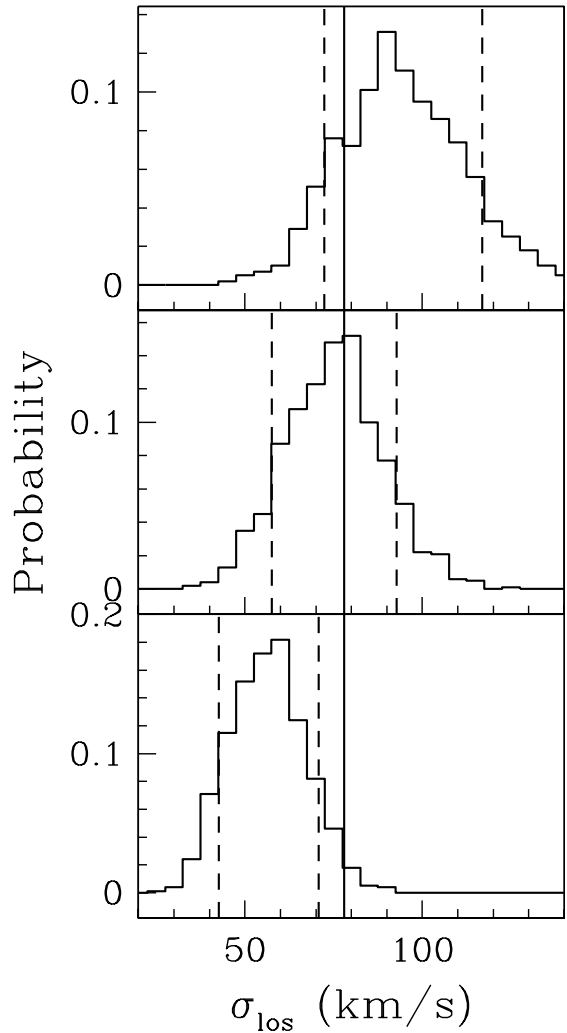}
\renewcommand{\baselinestretch}{0.5}
\caption{The probability distribution for \siglos\ measured
from the positions at $|Y|>6$ and $|X|>10$ for 1000 realizations of
various bulge models. The solid lines represent the value from the
data, \siglos\ = 79 km/s, and the dashed lines are the 10\% confidence
limits from the models.  Upper panel: a model bulge with \vrot\ =
160 km/s and \sigr\ = 90 km/s. Middle panel: a model bulge with
\vrot\ = 160 km/s and \sigr\ = 70 km/s.  Lower panel: a model bulge
with \vrot\ = 160 km/s and \sigr\ = 50 km/s.
\label{bulgedisp}}
\end{figure}

\begin{figure}
\includegraphics[scale=0.8]{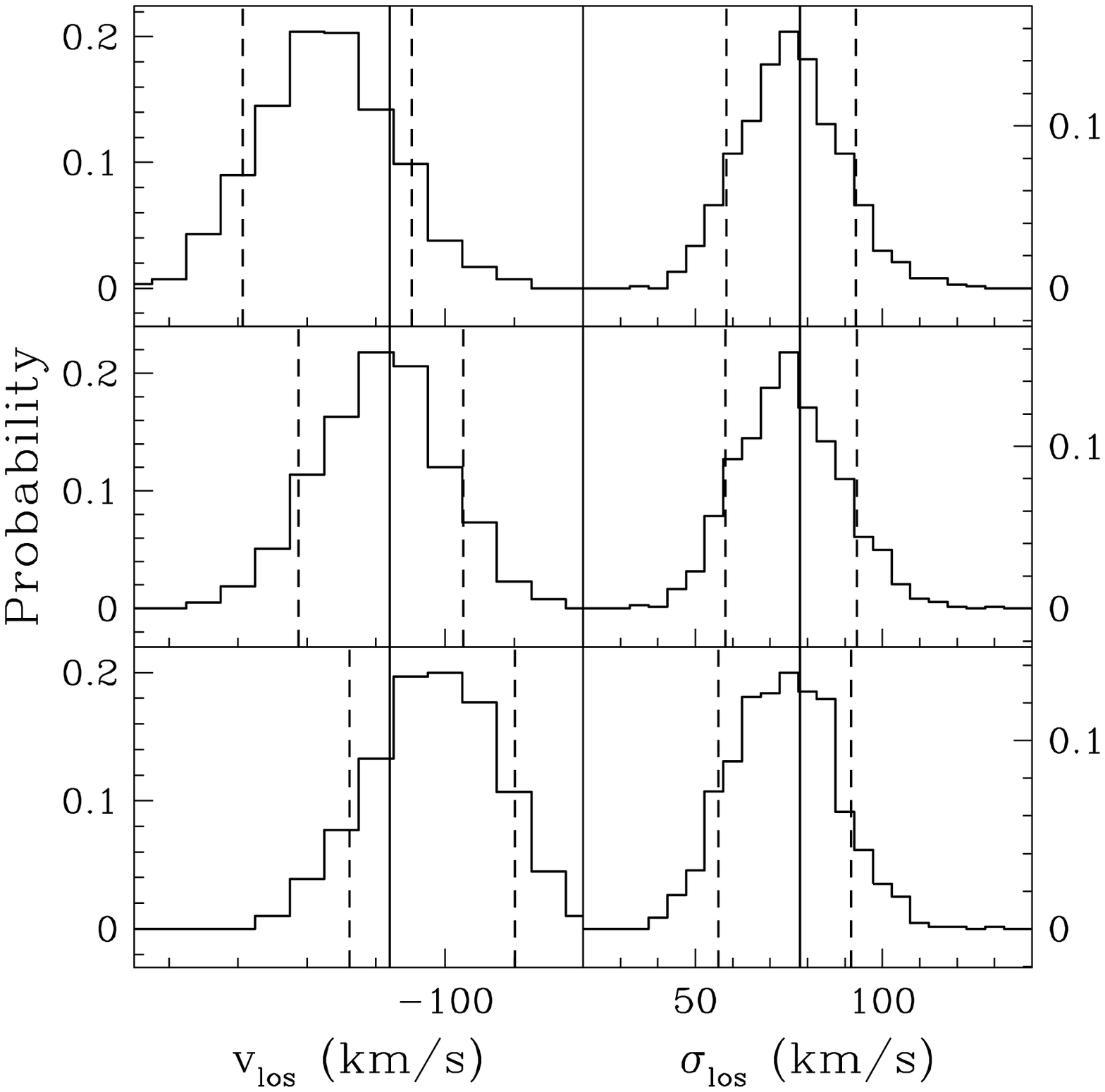}
\renewcommand{\baselinestretch}{0.5}
\caption{The probability distribution for the mean \vlos\ and for \siglos\
measured from the positions at $|Y|>6$ and $|X|>10$ for 1000
realizations of various bulge models. The solid lines represent the
values from the data, \vlos\ = 116 km/s and \siglos\ = 79 km/s, and
the dashed lines are the 10\% confidence limits from the models.
Upper panel: a model bulge with \vrot\ = 170 km/s and \siglos = 70
km/s. Middle panel: a model bulge with \vrot\ = 150 km/s and \sigr\
= 70 km/s.  Lower panel: a model bulge with \vrot\ = 130 km/s and
\sigr\ = 70 km/s.
\label{bulgemods}}
\end{figure}

We found from our models that the line-of-sight dispersion is
independent of the rotation velocity in the range of interest, and
that \sigr\ = 50 -- 90 km/s bracketed the range of reasonable values
(Figure \ref{bulgedisp}).  For those values, \vrot\ = 130 to 170 km/s
matched \vlos\ (Figure \ref{bulgemods}).  Our final best estimate is
\vrot\ = 150 $\pm$ 20 km/s and \sigr\ = $70 \pm 20$ km/s, where the
uncertainties cover the range of model parameters that led to a match
with the measured values of \vlos\ and \siglos.  In the lower right
panel of Figure \ref{realveldist}, we have plotted the range of
velocities generated for each PN position in the 1000 model runs of
this best model.  Only one outer PN is never fit by the bulge model,
in contrast to the disk models.  The dark line is the average rotation
curve for all the PN with $|Y|>$6, and the lighter lines are the model
10\% confidence limits for these PN.  Most fall within the limits --
their velocities are consistent with our bulge model.

Interestingly, \citet{reitzel04} find evidence of a population with
similar rotation in a field 34 kpc along the major axis.  Roughly
two-thirds of their sample of 23 M31 red giants have velocities
relative to M31 of 150 km/s, in good agreement with our model-based
estimate for the major axis rotation for the outer bulge.  However,
the dispersion is low, only 27 km/s, and \citet{reitzel04} reason that
these giants may belong to the disk, or cold substructure in the halo.
More velocities covering a wider area will be needed to disentangle
these populations.

For the outer PN, the model suggests \voversig\ = 150/70 =
2.1\footnote{We use \voversig\ in the same sense as \citet{rix99}: a
simple quotient of the line-of-sight velocity and velocity dispersion
at a given position}, well within the range covered by the
low-luminosity ellipticals \citep{rix99}.  If we accept that these PN
belong to the M31 bulge based on the surface brightness profile, then
the velocities imply that M31's bulge shares kinematic as well as
photometric properties with low luminosity ellipticals.  This
strengthens the evolutionary connection between these types of
galaxies, and may provide an important clue to understanding the
formation of large spirals like M31.

\subsection{The Halo}

Regardless of how we interpret the large rotation velocity of the
outer PN, it seems clear that most do not belong to a kinematically
hot, canonical halo.  Are there any PN in our sample which could?
Without recourse to more detailed models than justified by the data,
the best we can do is identify PN which cannot belong to either the
disk or the bulge.  At all $Y$, a few objects can be found that have
velocities which are inconsistent with either of these populations,
and their coordinates and velocities are listed in Table 6.  As a
group, their spectra do not systematically differ from the other PN.

\begin{figure}
\includegraphics[scale=0.8]{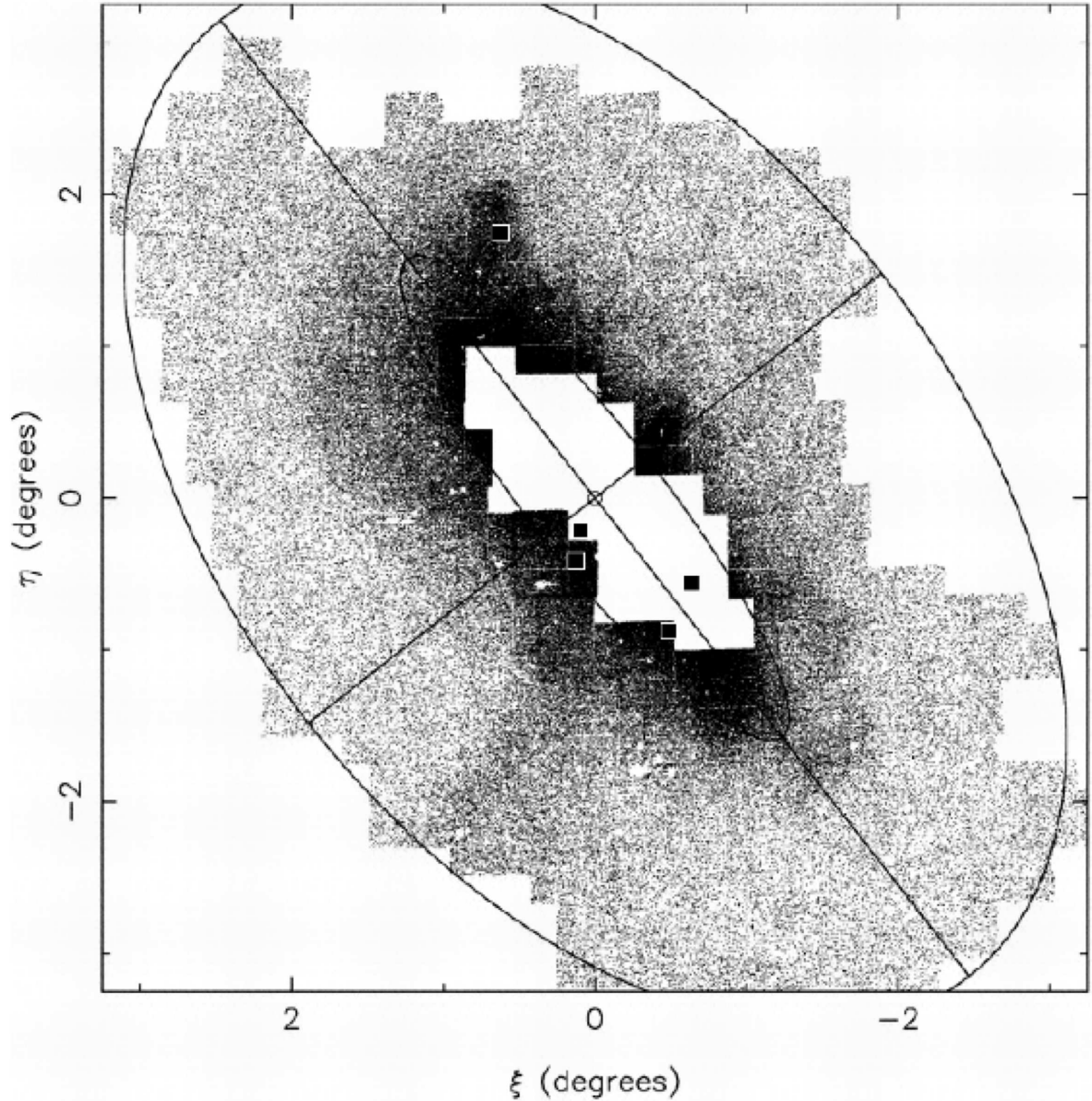}
\renewcommand{\baselinestretch}{0.5}
\caption{Figure 2 from \citealt{ferguson02}, showing M31's halo and
tidal streams, with the positions of our potential halo objects marked
(red in the electronic edition).  All 5 fall on or near possible tidal
streams, and three are counter-rotating with respect to the disk at
their position.  The two halo candidates near the center have
velocities within 40 km/s of each other, and have velocities relative
to M31 which are at least 3 times the central bulge dispersion.
Orbits suggested by \citet{merrett03} and \citet{ibata04} for the
Southern Stream \citep{ibata01,ferguson02} project near the position
of these PN.
\label{streampn}}
\end{figure}

The suprising conclusion we draw from the PN velocities and Figure
\ref{streampn}, where we have overplotted their positions on Figure 2
from \citet{ferguson02}, is that all five halo PN probably belong to
tidal streams in M31's outer halo.  One is projected on the Northern
spur, and is counter-rotating with respect to the disk there.  Two are
projected near the major axis at $X=-10$ kpc; their velocities are
similar to M32's systemic velocity and they could be debris from that
galaxy. All three of these halo PN have velocities in the ``forbidden"
quadrants.

The remaining two halo PN are located near the center of the galaxy.
Their velocities are within 40 km/s of each other ($v_{helio}\sim -600$
km/s) and are at least 3 times the central bulge dispersion (100
km/s).  Orbits suggested by \citet{merrett03} and \citet{ibata04} for
the Southern Stream \citep{ibata01,ferguson02} project near the
position of these PN.  Their velocities follow the gradient found by
\citet{ibata04}, implying that these PN could belong to the Southern
Stream.  Finding PN in low surface brightness streams is somewhat
unexpected, as PN are intrinsically rare objects (only
$\sim$30/$10^9$\lsolar; Ciardullo et al. 1989).  Velocities in the
streams will provide constraints on the orbits of their progenitors,
and deeper observations are planned to detect more stream PN.  We will
revisit these possible streams in a later paper.

We have yet to find the M31 analog to the hot component of the Milky
Way inner halo.  If the PN production rate per unit luminosity is the
same as for other M31 populations, we would expect to see of order
20--30 PN if we covered the entire halo of M31 at the depth of our
survey and its luminosity is of order $10^9$ \lsolar, the approximate
luminosity of the Milky Way halo \citep{morrison93}.  Depending on the
density distribution of the halo stars, we have surveyed one quarter
or less of the halo, so very roughly 5-7 PN would be expected in our
sample from such a halo.  Yet all 5 of our halo PN candidates belong
to kinematically cold features in the halo.

On the surface, this suggests that the M31 halo does not have a
kinematically hot component analogous to the inner Milky Way
halo. However, if the hot component of the M31 halo is as old and
metal-poor as the Milky Way halo, the PN production rate could be
lower by a factor of 5 or more \citep{magrini03}.  This would be
consistent with a total lack of hot halo objects in our sample.
Studies in search of these objects must reach farther down the PN
luminosity function by several magnitudes.

Spectroscopic surveys of red giant stars, because of the larger sample
size, may be needed to find the hot halo, especially if the PN
production rate is significantly lower for old populations.
\citet{reitzel02} found a high dispersion of 150 km/s in a sample of
M31 red giants in a field 19 kpc along the minor axis.  This field is
further out than our PN sample and may have reached, at last, a region
where the halo dominates.

\section{Discussion}

\subsection{Bulge vs. Halo}

The extent of M31's bulge has been a topic of debate in the
literature.  Some previous studies of M31 stellar populations
suggested that fields as far out as 20 kpc were still dominated by the
bulge (\citealt{mould86b}; \citetalias{pat01}; \citealt{freeman96})
primarily because of two important findings: first, the higher
metallicity of the stars compared to the Milky Way halo at comparable
radius (Durrell et al. 1994, 2001; \citealt{holland96}), and second,
its \rq\ surface brightness profile which extends smoothly out to at
least 20 kpc on the minor axis \citepalias{pvdb}.  We add to this
evidence the discovery that PN at $|Y|>$6 kpc, in regions typically
associated with the halo, have a \vlos\ = 116 km/s and
\siglos\ = 79 km/s.  This rotation is very different from the behavior 
of the Milky Way halo populations at similar radii and suggests that
the bulge light dominates even out to 20 kpc on the minor axis.

M31 globular clusters also differ significantly from their Milky Way
counterparts.  As a system, the clusters have a large rotation, $\sim
150$ km/s on average\citep{perrett02}.  This holds for both the
metal-rich and the metal-poor subsets, implying that neither have
clear ``halo'' kinematics, although the precise velocity dataset of
\citet{perrett02} was confined to inner regions and may select against
halo objects. \citet{morrison03a} demonstrate that $\sim50$\% of M31
clusters projected on the disk actually have thin disk kinematics.  No
analogous population is known among the Milky Way globular clusters.
In contrast, studies of the Milky Way globular clusters have found
that metal-poor clusters belong to the halo \citep{zinn85}, while the
central {\it metal-rich} clusters have either bulge or thick disk
kinematics \citep{{frenk82},{minniti95},{cote99},{armandroff89}}.

If canonical M31 halo regions are actually dominated by the bulge,
then further comparisons with the Milky Way {\it halo} are misleading.
Instead, these M31 halo regions provide a baseline for comparison to
more distant early type spiral bulges.  How does a bulge with moderate
rotational support form?  In major mergers, angular momentum can
migrate outwards in the dark halo as it is assembled \citep{quinn88}.
Thus, a bulge which formed in the very early stages of galaxy
formation simultaneously with the dark halo assembly would have the
\rq\ law profile characteristic of violent relaxation as well as
increased rotation at large radii.  PN studies of the outer regions of
some large ellipticals show this expected increase in rotation with
radius \citep[e.g.][]{hui93}.

Dissipationless mergers of disk galaxies of unequal mass (ratio 3:1)
can also lead to disky remnants with more rotational support in the
inner regions and \rq\ law profiles \citep{naab99,bendo00} as the
larger disk partly survives the interaction.  We find that the
kinematics of M31's outer bulge is in rough agreement with the
low-luminosity elliptical observations.  A comparison of M31's bulge
kinematics with these observations is complicated because velocities
for the ellipticals were measured along the major axis.  Nonetheless,
we find that \voversig\ = 2.1 by 5$R_e$, based on the outer bulge PN
kinematics in M31 (assuming $R_e$=2.4 kpc on the major axis;
\citetalias{pvdb}), similar to values for the low luminosity
ellipticals, suggesting an evolutionary kinship.

\subsection{The Elusive Halo Remains Elusive}

In our entire survey, we only find a handful of PN in the combined
sample with halo-like kinematics.  Ideally, these are the sample we
should compare to the Milky Way halo; however, our PN spectra are not
suitable for abundances, and tell us nothing about ages.  In
principle, they offer the possibility of comparing the luminosities of
the M31 and Milky Way halos, because the PN are direct tracers of the
underlying stars.  However, this is a difficult comparison to make for
several reasons.  First, uncertainties in the PN production rate make
it difficult to translate the number of PN to stellar luminosity.
Second, hot halo objects can have any velocity, and confusion with the
disk and bulge prevent us from unambiguously identifying those not in
the forbidden quadrants.  Third, the Milky Way halo luminosity is only
approximately known, and depends strongly on the density of the
little-studied inner halo, because the halo density distribution is
extremely centrally concentrated, with a globular cluster density
proportional to \Large $r^{-3.5}$ \normalsize \citep{zinn85}.
	   
Perhaps most important, a generalization from our survey to the entire
halo is not appropriate if M31's halo is clumpy. In fact, we know this
to be an important issue, as substructure in field star counts has
been detected in M31's halo \citep{ibata01,ferguson02}.  The
clumpiness of the Milky Way halo is now accepted, based on the
disruption of the Sgr dSph \citep{ibata94,ivezic00,yanny00,rdp} and on
studies of velocity substructures among the metal-poor halo giants in
the solar neighborhood \citep{helmi99}.

The PN observations thus far tell us more about the bulge than the
halo, as do other studies of the stellar populations in these fields.
The results of such studies can now be associated with the bulge; for
example, estimates of a suprisingly high halo luminosity 
\citepalias{reitzel98}, an intermediate-age component \citep{brown03}, 
and moderate flattening (\citetalias{pvdb}; \citealt{ferguson00}).

\section{Summary}

Our search for M31's halo has led to a picture of that galaxy very
different from that of the Milky Way.  PN at large distances from the
M31 major axis have significant rotation and small velocity
dispersion, in contrast to the large velocity dispersion of the Milky
Way halo.  We conclude that these objects do not belong to the halo,
but to the outer reaches of the bulge.

If this is the case, then among other characteristics, M31's bulge
shares the rotation properties of low-luminosity ellipticals; namely,
that the amount of rotational support increases with radius.  This
suggests an evolutionary kinship with low-luminosity ellipticals,
which are believed to have formed via hierarchical merging of objects
of nearly equal mass, in contrast to the Milky Way halo which formed
from the on-going accretion of very small satellites.

Any kinematically hot halo in M31 is represented by a very small
number of objects in our sample.  Yet given the rarity of PN, even
these few could imply a halo luminosity comparable to that of the
Milky Way, if they belong to a smooth inner component.  However, their
projection on M31 suggest they may instead belong to tidal streams in
a clumpy halo.  We are not claiming that that M31 does not have a
Milky-Way like halo. Rather we support, as some earlier researchers
have surmised, that the halo is dominated by the bulge out to 20 kpc
and farther; preliminary star counts of red giants in a field at 30
kpc reveal a similar metallicity distribution function to that of
nearer regions (P. Durrell 2003, private communication).

Sorting out M31's evolutionary history will require more data, and PN
provide the means.  With a larger sample of PN velocities, we can
clarify the distinction between bulge and halo by connecting the
kinematics to the spatial distribution.  Most important, abundances
can be derived for our sample of halo objects.  This is a great
advantage of PN; they can ultimately provide a large number of
abundances \citep[e.g.][]{jacoby99}, where this is costly in telescope
time for field stars \citep{reitzel02}.  The ability to tie abundance
to kinematics is a connection which has been critical to our
understanding of the stellar populations of the Milky Way.

\section{Acknowledgements}

We thank Rene Walterbos for providing the surface photometry of M31
and the image used in Figure \ref{image}; Annette Ferguson for the use
of Figure 2 from \citet{ferguson02}; John Feldmeier and Chris Mihos
for many useful discussions.  H.L.M acknowledges the support of
NSF CAREER grant AST-9624542.  D.H.K. is supported by an NSF Astronomy
and Astrophysics Postdoctoral Fellowship under award AST-0104455.

\begin{deluxetable}{lcccc} 
\tablecolumns{10} 
\tablewidth{0pc} 
\tablecaption{Coordinates and heliocentric velocities for PN candidates in M31.} 
\tablehead{\colhead{ID}&\colhead{RA (2000)}&\colhead{Dec (2000)}&\colhead{$v_{helio}$}&\colhead{$v_{err}$ }}
\startdata 
HKPN  1& 0:38:41.29& 39:47:40.1& --532.3&   4.0 \\
HKPN  2& 0:38:44.19& 40:17:58.9& --452.8&   5.1 \\
HKPN  3& 0:38:54.97& 40:10:16.2& --362.1&   1.9 \\
HKPN  4& 0:38:56.63& 39:47:14.2& --425.1&  12.6 \\
HKPN  5& 0:39:02.52& 40:22:50.5& --511.2&   3.1 \\
HKPN  6& 0:39:03.40& 39:45:58.4& --484.0&   3.4 \\
HKPN  7& 0:39:03.76& 39:53:29.0& --477.9&   3.5 \\
HKPN  8& 0:39:04.46& 40:34:30.3& --512.3&   3.0 \\
HKPN  9& 0:39:05.33& 40:41:46.0& --378.6&   5.3 \\
HKPN 10& 0:39:06.60& 40:14:59.8& --548.2&   4.3 \\
HKPN 11& 0:39:09.32& 40:11:21.0& --428.2&   2.5 \\
HKPN 12& 0:39:14.80& 40:24:28.6& --325.5&   6.4 \\
HKPN 13& 0:39:15.00& 40:26:34.5& --522.1&   5.9 \\
HKPN 14& 0:39:15.80& 40:12:38.5& --470.1&   4.5 \\
HKPN 15& 0:39:15.84& 40:12:39.3& --482.9&   5.3 \\
HKPN 16& 0:39:16.29& 40:22:12.8& --526.1&   5.6 \\
HKPN 17& 0:39:18.51& 40:09:19.0& --376.3&   4.1 \\
HKPN 18& 0:39:24.53& 40:07:02.4& --434.4&   2.7 \\
HKPN 19& 0:39:26.12& 40:44:26.0& --373.1&  27.7 \\
HKPN 20& 0:39:27.26& 40:29:03.6& --532.4&  12.6 \\
HKPN 21& 0:39:31.62& 40:14:19.9& --510.4&   3.6 \\
HKPN 22& 0:39:36.66& 40:17:51.3& --503.8&   3.4 \\
HKPN 23& 0:39:38.40& 39:21:45.6& --394.5&   8.8 \\
HKPN 24& 0:39:40.38& 40:13:24.7& --454.4&   3.1 \\
HKPN 25& 0:39:42.80& 40:22:49.2& --490.9&   8.0 \\
HKPN 26& 0:39:43.99& 40:35:42.0& --465.7&  12.1 \\
HKPN 27& 0:39:44.64& 40:19:59.2& --500.1&   3.1 \\
HKPN 28& 0:39:44.79& 40:12:04.2& --472.0&   6.5 \\
HKPN 29& 0:39:45.40& 40:31:42.6& --538.9&   4.6 \\
HKPN 30& 0:39:46.14& 40:11:01.7& --416.1&   2.1 \\
HKPN 31& 0:39:46.71& 39:38:42.6& --573.1&   3.2 \\
HKPN 32& 0:39:49.08& 40:42:39.5& --168.8&  28.6 \\
HKPN 33& 0:39:49.62& 40:21:25.1& --396.2&   4.0 \\
HKPN 34& 0:39:52.56& 40:09:32.4& --481.1&   6.4 \\
HKPN 35& 0:39:53.93& 40:31:56.5& --442.7&  24.6 \\
HKPN 36& 0:39:54.23& 39:51:00.0& --486.0&   3.3 \\
HKPN 37& 0:40:01.81& 40:37:28.7& --552.4&  10.6 \\
HKPN 38& 0:40:07.65& 40:28:34.1& --457.4&   5.3 \\
HKPN 39& 0:40:12.82& 40:13:02.4& --387.3&   4.3 \\
HKPN 40& 0:40:13.97& 40:10:26.4& --504.0&   2.3 \\
HKPN 41& 0:40:14.24& 40:23:23.9& --126.1&   2.5 \\
HKPN 42& 0:40:14.51& 40:38:50.7& --513.4&  12.2 \\
HKPN 43& 0:40:15.29& 40:24:13.6& --472.9&   4.4 \\
HKPN 44& 0:40:16.20& 40:09:59.7& --447.9&   6.9 \\
HKPN 45& 0:40:20.85& 40:28:08.9& --506.0&   2.9 \\
HKPN 46& 0:40:26.61& 40:27:59.3& --472.2&  26.9 \\
HKPN 47& 0:40:30.80& 40:36:54.4& --482.6&   2.6 \\
HKPN 48& 0:40:33.70& 40:13:46.8& --295.0&   4.7 \\
HKPN 49& 0:40:37.96& 40:13:35.5& --578.9&   3.3 \\
HKPN 50& 0:40:43.39& 39:57:02.3& --423.4&   2.7 \\
HKPN 51& 0:40:44.02& 40:25:08.7& --528.2&  36.7 \\
HKPN 52& 0:40:44.24& 40:27:11.2& --485.8&   6.1 \\
HKPN 53& 0:40:45.33& 40:41:30.3& --524.9&   5.1 \\
HKPN 54& 0:40:46.82& 39:51:48.5& --418.1&   6.9 \\
HKPN 55& 0:40:47.30& 40:28:25.6& --479.4&   7.7 \\
HKPN 56& 0:40:49.58& 40:39:46.8& --416.1&  18.7 \\
HKPN 57& 0:40:56.88& 40:20:09.0& --560.5&   3.9 \\
HKPN 58& 0:40:57.23& 40:36:07.8& --498.2&   8.6 \\
HKPN 59& 0:41:00.28& 40:28:50.0& --425.2&   2.4 \\
HKPN 60& 0:41:01.86& 40:24:20.3& --421.1&  23.5 \\
HKPN 61& 0:41:03.72& 40:41:13.6& --556.1&   6.0 \\
HKPN 62& 0:41:12.34& 40:35:58.2& --528.4&  10.0 \\
HKPN 63& 0:41:13.50& 40:42:15.8& --526.7&  14.0 \\
HKPN 64& 0:41:20.40& 39:28:10.6& --274.0&  13.8 \\
HKPN 65& 0:41:26.34& 39:59:48.4& --448.2&  17.7 \\
HKPN 66& 0:41:26.48& 40:32:45.4& --399.6&   4.0 \\
HKPN 67& 0:41:26.84& 40:43:16.2& --394.0&  14.1 \\
HKPN 68& 0:41:33.67& 40:20:39.5& --396.6&   3.8 \\
HKPN 69& 0:41:38.18& 39:54:59.8& --413.5&   2.3 \\
HKPN 70& 0:42:03.33& 40:31:39.8& --539.0&   4.0 \\
HKPN 71& 0:42:04.07& 40:29:37.7& --513.8&  26.7 \\
HKPN 72& 0:42:05.09& 40:43:16.3& --449.4&  44.3 \\
HKPN 73& 0:42:19.12& 40:57:09.0& --281.5&   8.7 \\
HKPN 74& 0:42:21.19& 40:33:49.9& --379.7&   4.2 \\
HKPN 75& 0:42:22.82& 40:43:06.6& --509.1&   6.9 \\
HKPN 76& 0:42:29.75& 41:03:30.4& --451.3&   5.4 \\
HKPN 77& 0:42:32.35& 40:42:26.2& --362.2&   8.6 \\
HKPN 78& 0:42:37.21& 41:04:19.4& --461.1&   8.0 \\
HKPN 79& 0:42:44.49& 40:17:31.0& --458.9&   3.7 \\
HKPN 80& 0:42:46.60& 40:26:15.2& --462.8&   3.9 \\
HKPN 81& 0:42:49.84& 40:51:10.6& --505.7&  10.0 \\
HKPN 82& 0:42:50.58& 40:04:48.5& --477.6&   4.5 \\
HKPN 83& 0:42:52.05& 40:29:03.1& --261.1&  20.1 \\
HKPN 84& 0:42:52.11& 40:29:03.9& --294.1&   2.6 \\
HKPN 85& 0:42:55.60& 41:03:40.9& --599.1&   5.6 \\
HKPN 86& 0:42:56.02& 40:51:12.5& --638.3&   5.3 \\
HKPN 87& 0:42:56.18& 40:35:40.4& --534.8&   8.6 \\
HKPN 88& 0:42:57.81& 41:12:43.4& --199.2&  18.0 \\
HKPN 89& 0:42:58.82& 41:22:13.2& --142.5&   4.6 \\
HKPN 90& 0:43:03.15& 40:57:18.5& --367.4&  11.0 \\
HKPN 91& 0:43:05.26& 40:53:34.1& --532.3&   4.8 \\
HKPN 92& 0:43:06.16& 41:10:56.0& --337.9&   3.9 \\
HKPN 93& 0:43:13.72& 40:52:32.4& --381.5&   6.0 \\
HKPN 94& 0:43:15.18& 40:51:12.7& --500.3&   6.2 \\
HKPN 95& 0:43:19.00& 40:58:02.1& --452.0&   4.3 \\
HKPN 96& 0:43:20.61& 40:55:32.4& --309.9&  11.2 \\
HKPN 97& 0:43:20.63& 40:57:44.5& --502.8&  12.0 \\
HKPN 98& 0:43:21.64& 40:59:12.7& --388.8&  28.8 \\
HKPN 99& 0:43:23.00& 41:15:40.4& --251.3&  10.0 \\
HKPN 100& 0:43:28.32& 40:51:19.9& --433.9&  25.2 \\
HKPN 101& 0:43:36.79& 41:03:33.4& --260.3&   7.4 \\
HKPN 102& 0:43:36.80& 40:37:38.6& --323.9&  12.4 \\
HKPN 103& 0:43:44.52& 40:59:59.7& --274.7&   4.3 \\
HKPN 104& 0:44:04.65& 41:12:08.9& --293.0&   4.7 \\
HKPN 105& 0:44:04.94& 41:09:18.6& --212.8&  13.7 \\
HKPN 106& 0:44:07.49& 41:10:09.7& --270.6&   6.4 \\
HKPN 107& 0:44:08.01& 40:03:13.6& --391.4&   3.8 \\
HKPN 108& 0:44:11.42& 41:01:02.2& --348.6&   6.0 \\
HKPN 109& 0:44:11.92& 41:14:00.0& --232.4&   4.5 \\
HKPN 110& 0:44:15.88& 41:22:10.4& --251.2&   5.0 \\
HKPN 111& 0:44:21.04& 41:15:41.1& --281.8&   4.2 \\
HKPN 112& 0:44:25.50& 40:14:36.6& --428.5&   8.0 \\
HKPN 113& 0:44:25.98& 40:15:45.7& --350.4&   3.2 \\
HKPN 114& 0:44:30.33& 40:07:29.5& --425.7&   2.3 \\
HKPN 115& 0:44:37.91& 41:14:14.4& --219.3&  15.0 \\
HKPN 116& 0:44:37.97& 41:07:02.9& --335.9&  11.7 \\
HKPN 117& 0:44:42.22& 41:08:01.8& --243.5&   4.9 \\
HKPN 118& 0:44:42.77& 40:40:05.7& --383.8&   4.7 \\
HKPN 119& 0:44:43.80& 41:14:28.0& --315.7&   5.9 \\
HKPN 120& 0:44:50.69& 41:17:22.3& --288.6&   5.3 \\
HKPN 121& 0:44:53.16& 41:15:58.0& --306.6&  28.5 \\
HKPN 122& 0:44:54.43& 40:52:45.4& --453.7&   3.8 \\
HKPN 123& 0:44:54.98& 41:09:53.2& --321.1&   5.1 \\
HKPN 124& 0:44:56.30& 41:20:25.4& --296.6&   4.5 \\
HKPN 125& 0:45:05.05& 41:12:52.5& --201.0&  10.9 \\
HKPN 126& 0:45:05.30& 41:09:38.6& --278.5&   6.3 \\
HKPN 127& 0:45:07.14& 40:39:33.4& --285.8&   7.7 \\
HKPN 128& 0:45:28.36& 41:04:54.2& --400.5&  16.8 \\
HKPN 129& 0:45:35.32& 40:59:57.2& --205.1&   9.4 \\
HKPN 130& 0:45:38.13& 40:59:19.5& --279.0&  10.9 \\
HKPN 131& 0:47:00.98& 40:49:23.4& --350.5&   3.1 \\
HKPN 132& 0:48:22.14& 40:45:41.9& --320.0&   3.4 \\
HKPN 133& 0:48:24.78& 41:08:11.1& --353.2&   3.3 \\
HKPN 134& 0:48:27.23& 39:55:34.0& --186.8&   2.4 \\
HKPN 135& 0:49:28.30& 40:59:54.0& --242.6&   3.4 \\
\enddata 
\end{deluxetable}

\clearpage

\begin{deluxetable}{lcccc} 
\tablecolumns{5} 
\tablewidth{0pc} 
\tablecaption{PN candidates in M32.} 
\tablehead{ 
\colhead{ID}    &\colhead{$\alpha$(2000)} &   \colhead{$\delta$(2000)}   & 
\colhead{$v_{helio}$} &\colhead{$v_{err}$}}
\startdata 
1& 0:42:23.24& 40:46:28.7& --182.4&   8.4 \\
2& 0:42:28.93& 40:44:39.0& --158.3&   6.6 \\
3& 0:42:40.10& 40:49:41.3& --160.5&   3.6 \\
4& 0:42:40.24& 40:51:03.0& --149.6&   2.3 \\
5& 0:42:50.63& 40:45:28.4& --183.7&   2.3 \\
6& 0:42:53.02& 40:48:59.8& --213.5&   2.0 \\
7& 0:42:56.97& 40:51:01.8& --169.1&  24.8 \\
8& 0:43:02.04& 40:49:30.2& --154.6&   4.7 \\
\enddata 
\end{deluxetable}

\clearpage

\begin{deluxetable}{lcccc} 
\tablecolumns{5} 
\tablewidth{0pc} 
\tablecaption{HII Regions in And IV} 
\tablehead{ 
\colhead{ID}    &\colhead{$\alpha$(2000)} &   \colhead{$\delta$(2000)}   & 
\colhead{$v_{helio}$} &\colhead{$v_{err}$}}
\startdata 
   6& 0:42:30.57& 40:34:46.9&  270.2&  39.2 \\
   3& 0:42:32.17& 40:33:58.7&  233.0&   3.4 \\
   4& 0:42:31.73& 40:34:11.2&  249.5&   5.2 \\
\enddata 
\tablenotetext{a}{ID numbers correspond to those given in Ferguson et al. (2000).}
\end{deluxetable}

\clearpage

\begin{table}
\caption{Kinematic Models} 
\centering
\vspace{.2in}
\begin{tabular} {lccccccc} 
\hline\hline\\
Model & $h_r$ & $h_z$ & $v_{rot}$  & Density & $\sigma_r$ & $\sigma_{\phi},\sigma_z$ \\
 &kpc & kpc & km/s &Distribution &km/s &km/s \\
\\
\hline\\
thin & 5.9 & 0.3 &  230 & $L(R,z) = e^{(-R/h_r)}e^{(-z/h_z)} $ & $ \sigma_r = \sqrt{h_z} v_{circ} e^{-R/2h_r} $ & $ (\sigma_{\phi},\sigma_z)=(\frac{\sigma_r}{\sqrt{2}},\frac{\sigma_r}{2})$ \\
thick & 5.9 &  1.0 & 230 & $ L(R,z) = e^{(-R/h_r)}e^{(-z/h_z)} $ & $ \sigma_r = \sqrt{h_z} v_{circ} e^{-R/2h_r} $ & $ (\sigma_{\phi},\sigma_z)=(\frac{\sigma_r}{\sqrt{2}},\frac{\sigma_r}{2})$  \\
bulge & 2.4 &... & 150 & $ L(R,z) = e^{-7.67((-R/R_e)^{1/4}-1)} $ & $ \sigma_r = constant $ & $ \sigma_{\phi} = \sigma_z = \sigma_r$ \\
\\
\hline\\
\end{tabular}
\end{table}

\begin{deluxetable}{lcccc} 
\tablecolumns{5} 
\tablewidth{0pc} 
\tablecaption{Milky Way Disk Properties.} 
\tablehead{\colhead{Population}&\colhead{$h_r$}&\colhead{$h_z$}&\colhead{$v_{circ}$}&\colhead{$\sigma_r$ }}
\startdata 
Milky Way thin & 3.5 & .3 & -220 & 30 \\
Milky Way thick & 4  & 1.0 & -220 & 65 \\
\enddata
\end{deluxetable}

\clearpage

\begin{deluxetable}{lcccc} 
\tablecolumns{5} 
\tablewidth{0pc} 
\tablecaption{Coordinates and heliocentric velocities for Halo PN candidates in M31.} 
\tablehead{\colhead{ID}&\colhead{RA (2000)}&\colhead{Dec (2000)}&\colhead{$v_{helio}$}&\colhead{$v_{err}$ }}
\startdata 
HKPN 86& 0:42:56.02& 40:51:12.5& --638.3&   5.3 \\
HKPN 85& 0:42:55.60& 41:03:40.9& --599.1&   5.6 \\
HKPN 32& 0:39:49.08& 40:42:39.5& --168.8&  28.6 \\
HKPN 41& 0:40:14.24& 40:23:23.9& --126.1&   2.5 \\
NF&  0:46:26.51 &  43:00:42.15 & --452.3&  .... \\
\enddata 
\tablecomments{The last object is from the study of \citealt{nolthenius87}.}
\end{deluxetable}

\end{document}